\documentclass[twocolumn]{aastex7}

\usepackage{amsmath}
\usepackage{wasysym}
\usepackage{savesym}
\savesymbol{tablenum}
\usepackage{siunitx}
\restoresymbol{SIX}{tablenum}
\usepackage{booktabs}
\usepackage{rotating}
\usepackage[T1]{fontenc}
\usepackage{blindtext}
\usepackage{tensor}
\usepackage{dsfont}

\newcommand{\upd}{\mathrm{d}}

\DeclareSIUnit\Mearth{M_\oplus}
\DeclareSIUnit\Mjupiter{M_J}
\DeclareSIUnit\Rearth{R_\oplus}
\DeclareSIUnit\Rjupiter{R_J}
\DeclareSIUnit\Mmoon{M_\textrm{\leftmoon}}
\DeclareSIUnit\JEM{J_{EM}}

\graphicspath{{./}{figures/}}

\received{20.03.2026}
\revised{18.05.2026}
\accepted{20.05.2026}
\submitjournal{APJ}

\begin{document}

\title{Smoothed Particle Hydrodynamics in \texttt{pkdgrav3} for Shock Physics Simulations. II. Shear Strength}

\author[orcid=0000-0001-9682-8563,gname='Thomas', sname='Meier']{Thomas Meier} 
\affiliation{Department of Astrophysics, University of Zurich, Winterthurerstrasse 190, CH-8057 Zurich, Switzerland}
\email[show]{thomas.meier5@uzh.ch}

\author[orcid=0000-0002-4535-3956,gname='Christian' ,sname='Reinhardt']{Christian Reinhardt}
\affiliation{Department of Astrophysics, University of Zurich, Winterthurerstrasse 190, CH-8057 Zurich, Switzerland}
\affiliation{Physics Institute, Space Research and Planetary Sciences, Center for Space and Habitability, University of Bern, Sidlerstrasse 5, CH-3012 Bern, Switzerland}
\email{christian.reinhardt@uzh.ch}

\author[orcid=0000-0002-1800-2974,gname='Martin', sname='Jutzi']{Martin Jutzi} 
\affiliation{Physics Institute, Space Research and Planetary Sciences, Center for Space and Habitability, University of Bern, Sidlerstrasse 5, CH-3012 Bern, Switzerland}
\email{martin.jutzi@unibe.ch}

\author[orcid=0000-0002-0757-5195,gname='Douglas', sname='Potter']{Douglas Potter} 
\affiliation{Department of Astrophysics, University of Zurich, Winterthurerstrasse 190, CH-8057 Zurich, Switzerland}
\email{joachimgerhard.stadel@uzh.ch}

\author[orcid=0000-0001-7565-8622,gname='Joachim', sname='Stadel']{Joachim Stadel} 
\affiliation{Department of Astrophysics, University of Zurich, Winterthurerstrasse 190, CH-8057 Zurich, Switzerland}
\email{joachimgerhard.stadel@uzh.ch}

\begin{abstract}
Material strength effects have been recently shown to be significant in giant impacts even at scales of planetary collisions. Despite this, their effects are often neglected in numerical giant impact simulations. We present an implementation of a basic strength model (pressure dependent shear strength) in the massively parallel smoothed particle hydrodynamics code \texttt{pkdgrav3}. The model includes elastic deviatoric stresses, plasticity with pressure-dependent yield strength, and thermal softening, and is fully integrated into the GPU-accelerated framework introduced in Paper~I, preserving its scalability and performance characteristics. We validate the implementation against laboratory experiments of granular cliff collapse and our simulation results are in excellent agreement. We then determine the catastrophic disruption threshold, $Q_{RD}^*$, over a wide mass range of the colliding bodies using simulations performed both with and without material strength. Consistent with prior work, we find that strength substantially increases $Q_{RD}^*$ in the low-mass regime, while convergence toward the fluid limit occurs only near $R_{C1} \sim \SI{e7}{\meter}$ ($\sim 0.7,M_\oplus$), well above the often assumed $\sim \SI{100}{\kilo\meter}$ size limit. Entropy production and remnant morphology likewise remain sensitive to rheology at intermediate masses. Performance measurements show that including strength introduces only modest computational overhead while maintaining favorable scaling, thereby enabling realistic solid mechanics in large-scale impact simulations.
\end{abstract}



\section{Introduction} 
Collisions are a fundamental driver of structure formation in the Solar System and exoplanetary systems \citep{safronovRelativeSizesLargest1969}. From the growth of planetesimals during early accretion to catastrophic disruption events during the formation and evolution of the asteroid belt and giant impacts during terrestrial planet formation, impact processes influence the mass distribution, internal structure, thermodynamic state, and morphology of solid bodies across a wide range of scales \citep{canupDynamicsLunarFormation2004,chauFormingMercuryGiant2018,reinhardtBifurcationHistoryUranus2020,reinhardtFormingIronrichPlanets2022,ballantyneInvestigatingFeasibilityImpactinduced2023,ballantyneSputnikPlanitiaImpactor2024,raducanPhysicalPropertiesAsteroid2024,meierOriginJupitersFuzzy2025,dentonCaptureAncientCharon2025,wimarssonDiverseShapesBinary2025,cambioniFormationAsteroid162026}. Numerical modeling, and in particular smoothed particle hydrodynamics (SPH) \citep{lucyNumericalApproachTesting1977,monaghanShockSimulationParticle1983}, has become a primary tool for studying these highly non-linear, energetic events \citep{benzImpactSimulationsFracture1994,canupOriginMoonGiant2001,jutziNumericalSimulationsImpacts2008,jutziModelingAsteroidCollisions2015}.

In many large-scale impact simulations, material strength is neglected under the assumption that self-gravity dominates the mechanical response of sufficiently massive bodies \citep{benzOriginMoonSingleimpact1986,cukMakingMoonFastSpinning2012,gendaResolutionDependenceDisruptive2015,chauFormingMercuryGiant2018,reinhardtBifurcationHistoryUranus2020,meierEOSResolutionConspiracy2021,kegerreisImmediateOriginMoon2022,timpeSystematicSurveyMoonforming2023,meierSystematicSurveyMoonforming2024,bussmannPossibilityGiantImpact2025}. This approximation is well justified in the limit of giant impacts between planetary-mass objects. However, the transition between strength-dominated and gravity-dominated regimes is gradual rather than abrupt, and the mass scale at which material strength becomes dynamically irrelevant in energetic collisions remains insufficiently quantified \citep{benzCatastrophicDisruptionsRevisited1999,jutziFragmentPropertiesCatastrophic2010,jutziModelingAsteroidCollisions2015,jutziSPHCalculationsAsteroid2015,emsenhuberSPHCalculationsMarsscale2018,ballantyneInvestigatingFeasibilityImpactinduced2023,ballantyneSputnikPlanitiaImpactor2024,dentonCaptureAncientCharon2025}. One way of estimating the scale at which self-gravity overcomes static shear strength in shaping equilibrium bodies is the ``potato radius'' \citep{lineweaverPotatoRadiusLower2010}, however, recent studies have shown that it does not indicate the scale at which strength ceases to influence transient, high-strain-rate collision outcomes. This is particularly evident in cratering events, where even relatively small bodies can exhibit strength-dominated responses if the impact energy is low, highlighting that both the target mass and the energy of the collision jointly determine the relative importance of material strength \citep{nakamuraImpactCrateringPorous2017,kiuchiImpactExperimentsGranular2023}. In particular, the disruption threshold $Q_{RD}^*$, fragment size distributions, entropy production, and remnant morphology may retain sensitivity to material rheology well above this scale \citep{benzCatastrophicDisruptionsRevisited1999,jutziFragmentPropertiesCatastrophic2010,emsenhuberSPHCalculationsMarsscale2018,ballantyneInvestigatingFeasibilityImpactinduced2023,dentonCaptureAncientCharon2025}.

Over the past decades, various strength and damage models have been incorporated into shock physics codes, including pressure-dependent shear strength, tensile fragmentation, cohesion, and porosity appropriate for geological materials \citep{druckerSoilMechanicsPlastic1952,gradyContinuumModellingExplosive1980,mcglaunCTHThreedimensionalShock1990,meloshDynamicFragmentationImpacts1992,benzSimulationsBrittleSolids1995,collinsModelingDamageDeformation2004,wunnemannStrainbasedPorosityModel2006,jutziNumericalSimulationsImpacts2008,maindlSPHbasedSimulationMultimaterial2013,michaelowenCompatiblyDifferencedTotal2014,jutziSPHCalculationsAsteroid2015,jutziModelingAsteroidCollisions2015,schaferSmoothParticleHydrodynamics2016,sugiuraHighresolutionSimulationsCatastrophic2020}. These developments have enabled increasingly realistic simulations of small-body collisions. At the same time, advances in high-performance computing have made it possible to perform impact simulations with billions of particles, resolving fine-scale structures and reducing numerical dissipation. Despite these advances, no study to date has combined fully realistic material physics, including strength, damage, porosity, and cohesion, with extreme computational scalability.

In Paper~I \citep{meierSmoothedParticleHydrodynamics2026}, we presented the SPH implementation in the massively parallel code \texttt{pkdgrav3}, demonstrating excellent performance and scaling characteristics on modern GPU-accelerated architectures. In this work, we extend that framework by incorporating a pressure-dependent shear strength model into the existing SPH framework. The implementation is fully integrated into the established data structures and parallelization strategy, preserving the GPU-centric design and hybrid MPI-threaded execution model.

We validate our strength model implementation against laboratory experiments and apply it to a range of collision scenarios to explore the role of material strength across different size scales. In addition, we assess the computational performance and scaling of the implementation on modern supercomputing systems. This work establishes a framework for incorporating realistic strength models into large-scale impact simulations, providing a foundation for future studies that include porosity, tensile fragmentation, and more advanced solid-mechanics models.

This paper is organized as follows. In Section~\ref{sec:Code_Description}, we describe the implementation of the shear strength model in \texttt{pkdgrav3}. Section~\ref{sec:Tests_with_strength} presents a series of tests to validate the code and illustrate the effects of material strength. In Section~\ref{sec:Performance}, we examine how the inclusion of strength influences computational performance. Finally, Section~\ref{sec:Conclusions} provides a summary of our findings and concluding remarks.

\section{Code Description}\label{sec:Code_Description}
The shear strength model is implemented within the existing smoothed particle hydrodynamics (SPH) framework of \texttt{pkdgrav3}, building directly on the SIMD- and GPU-optimized architecture described in Paper~I \citep{meierSmoothedParticleHydrodynamics2026}. Rather than introducing separate solver components, the strength model reuses the same data layout, neighbor interaction loops, and time-integration infrastructure as the hydrodynamic formulation. This approach allows the additional terms necessary for the material strength formulation to be evaluated alongside the standard SPH quantities with minimal computational overhead, while preserving the scalability, memory locality, and performance characteristics of the original implementation. As a result, simulations that include material strength retain the computational efficiency of the baseline SPH solver presented in Paper~I.

The source code of \texttt{pkdgrav3}, including both the SPH and shear strength implementations, is released under the GNU General Public License (version 3, GPLv3)\footnote{\url{https://www.gnu.org/licenses/gpl-3.0.en.html}} on Bitbucket\footnote{\url{https://bitbucket.org/dpotter/pkdgrav3}}, with the current version deposited to Zenodo: \href{https://zenodo.org/records/18754678}{10.5281/zenodo.18754678} \citep{potterDpotterPkdgrav3V3512026}. Documentation is available\footnote{\url{https://pkdgrav3.readthedocs.io/}}.

In this Section, we first summarize the SPH formulation presented in Paper~I, including the equations of motion and the specific implementation choices. We then outline the formulation of the shear strength model, followed by the treatment of plasticity, including the yield criterion and stress rescaling procedure, and the numerical time integration of the deviatoric stress and strain-rate tensors.

\subsection{Smoothed Particle Hydrodynamics}
The SPH implementation in \texttt{pkdgrav3} on which we build our shear strength implementation is described in detail in Paper~I \citep{meierSmoothedParticleHydrodynamics2026}. Here we briefly summarize the key aspects of the numerical scheme relevant to this study.

The density is computed via kernel summation as

\begin{equation}\label{eq:Density_estimate}
\rho_i = \sum_jm_jW_{ij}(h_i)\,,
\end{equation}

\noindent where $W_{ij}(h_x) = W(\left\vert\vec{r}_i-\vec{r}_j\right\vert,h_x)$ is the smoothing kernel and $h_i$ is the adaptive smoothing length. To obtain the smoothing length, the density and its derivative with respect to $h_i$ are computed and used in Newton-Raphson iterations to maintain a constant effective mass in the kernel \citep{springelCosmologicalSmoothedParticle2002,priceSmoothedParticleHydrodynamics2012,hopkinsGeneralClassLagrangian2013}:

\begin{equation}
M_{tot}^i=\int_{V_i}\rho_i \upd V \approx \frac{4}{3}\pi R_{kernel}^3(h_i)\rho_i\,.
\end{equation}

\noindent This procedure results in a smooth spatial variation of $h_i$ and enables the use of the $\nabla h$ correction term accounting for gradients of the smoothing length,

\begin{equation}
\Omega_i = 1 + \frac{h_i}{3\rho_i}\sum_jm_j\frac{\partial W_{ij}(h_i)}{\partial h_i}\,.
\end{equation}

The accelerations are then calculated as

\begin{equation}
\left.\frac{\upd \vec{v}_i}{\upd t}\right|_{H} =-\sum_jm_j\left[\frac{P_i}{\Omega_i\rho_i^2}\frac{\partial W_{ij}(h_i)}{\partial \vec{r}_i}+\frac{P_j}{\Omega_j\rho_j^2}\frac{\partial W_{ij}(h_j)}{\partial\vec{r}_i}\right]\,.
\end{equation}

In order to capture shocks, we use artificial viscosity in the original SPH form \citep{monaghanSmoothedParticleHydrodynamics1992}

\begin{equation}
\Pi_{ij} = \begin{cases}\frac{-\alpha \overline{c}_{ij}\mu_{ij}+\beta\mu_{ij}^2}{\overline{\rho}_{ij}}&\mbox{for } \vec{v}_{ij}\cdot\vec{r}_{ij}<0\\
0&\mbox{otherwise}\end{cases}\,,
\end{equation}

\noindent where

\begin{equation}
\mu_{ij}=\frac{\overline{h}_{ij}(\vec{v}_{ij}\cdot\vec{r}_{ij})}{\vec{r}_{ij}^2+\epsilon \overline{h}_{ij}^2}\,,
\end{equation}

\noindent where $\overline{c}_{ij}$, $\overline{\rho}_{ij}$ and $\overline{h}_{ij}$ are the averages of the respective values for particles $i$ and $j$. $\alpha$ and $\beta$ represent the shear and Von Neumann-Richtmyer viscosities respectively. The artificial viscosity generates contributions (denoted by the subscript $\Pi$) to both the accelerations

\begin{equation}
\left.\frac{\upd\vec{v}_i}{\upd t}\right|_{\Pi} = -\sum_jm_j\Pi_{ij}\frac{\partial \overline{W}_{ij}}{\partial\vec{r}_i}\,,
\end{equation}

\noindent and the internal energy derivatives

\begin{equation}\label{eq:AV_heating}
\left.\frac{\upd u_i}{\upd t}\right|_{\Pi}  = \frac{1}{2}\sum_jm_j\Pi_{ij}(\vec{v}_i-\vec{v}_j)\cdot\frac{\partial \overline{W}_{ij}}{\partial\vec{r}_i}\,,
\end{equation}

\noindent where we use the averaged kernel $\overline{W}_{ij} = 0.5(W_{ij}(h_i) + W_{ij}(h_j))$.

All simulations in this work use the standard kernel summation density estimate given in Equation~\eqref{eq:Density_estimate}. To improve the density estimation at material and vacuum interfaces, we employ the interface correction of \citet{ruiz-bonillaDealingDensityDiscontinuities2022} as described in Section~2.3.4 of Paper~I. In order to strictly enforce conservation of entropy in the absence of shocks, we employ the method presented by \citet{reinhardtNumericalAspectsGiant2017} and described in Section~2.3.5 of Paper~I to compute the adiabatic evolution of the internal energy. While these methods have been validated in previous work, their explicit validation in combination with material strength is not performed as part of the present study. However, for the tests presented below, we do not observe any unexpected behavior resulting from their combined use.

\subsection{Shear strength}\label{sec:Shear_strength}
In order to model solid materials, the standard smoothed particle hydrodynamics (SPH) formulation can be extended to include shear strength \citep{benzImpactSimulationsFracture1994,benzSimulationsBrittleSolids1995}. This is achieved by generalizing the scalar pressure $P$ to the full Cauchy stress tensor $\sigma^{\alpha\beta}$,

\begin{equation}
\sigma^{\alpha\beta} = -P\delta^{\alpha\beta}+S^{\alpha\beta}\,,
\end{equation}

\noindent where $\delta^{\alpha\beta}$ denotes the Kronecker delta. The first term, $-P\delta^{\alpha\beta}$, represents the isotropic (hydrostatic) contribution to the stress, while $S^{\alpha\beta}$ is the deviatoric stress tensor describing shear stresses within the material. By construction, the deviatoric stress tensor is symmetric and traceless. As a consequence, it contains only five independent components. Writing the stress tensor explicitly in matrix form highlights this decomposition:

\begin{align}
&\begin{pmatrix}
\sigma^{xx}&\sigma^{xy}&\sigma^{xz}\\\sigma^{yx}&\sigma^{yy}&\sigma^{yz}\\\sigma^{zx}&\sigma^{zy}&\sigma^{zz}\end{pmatrix}=\\
&-\begin{pmatrix}
		P&0&0\\0&P&0\\0&0&P
	\end{pmatrix}+\begin{pmatrix}
		S^{xx}&S^{xy}&S^{xz}\\S^{xy}&S^{yy}&S^{yz}\\S^{xz}&S^{yz}&-S^{xx}-S^{yy}
	\end{pmatrix}\,.
\end{align}

\noindent To simplify the notation, from here on, we use Einstein summation convention, with Greek indices indicating summation over spatial coordinates.

For an isotropic, linear elastic continuum, the deviatoric stress tensor is related to the strain tensor $\epsilon^{\alpha\beta}$ through

\begin{equation}
S^{\alpha\beta}=2\Gamma\left(\epsilon^{\alpha\beta}-\frac{1}{3}\epsilon^{\gamma\gamma}\delta^{\alpha\beta}\right)\,,
\end{equation}

\noindent where $\Gamma$ is the shear modulus, a material-dependent constant. Here summation over $\gamma$ is implied, i.e., $\epsilon^{\gamma\gamma} \equiv \mbox{tr}(\epsilon)$. The subtraction of this trace term ensures that $S^{\alpha\beta}$ remains traceless, consistent with its definition.

Since the material derivative of the above expression is not frame invariant, we instead adopt the Jaumann rate to obtain an objective stress rate, leading to

\begin{equation}
\dot{S}^{\alpha\beta}=2\Gamma\left(\dot{\epsilon}^{\alpha\beta}-\frac{1}{3}\dot{\epsilon}^{\gamma\gamma}\delta^{\alpha\beta}\right)+S^{\alpha\gamma}R^{\gamma\beta}+S^{\beta\gamma}R^{\gamma\alpha}\label{eq:Deviatoric_Stress_derivative}\,,
\end{equation}

\noindent where $\dot{\epsilon}^{\alpha\beta}$ is the strain-rate tensor and $R^{\alpha\beta}$ is the rotation-rate tensor. These are defined in terms of the velocity field $v^{\alpha}$ as

\begin{align}
\dot{\epsilon}^{\alpha\beta}&=\frac{1}{2}\left(\frac{\partial v^\alpha}{\partial x^\beta}+\frac{\partial v^\beta}{\partial x^\alpha}\right)\label{eq:Strain_rate_tensor}\,,\\
R^{\alpha\beta}&=\frac{1}{2}\left(\frac{\partial v^\alpha}{\partial x^\beta}-\frac{\partial v^\beta}{\partial x^\alpha}\right)\label{eq:Rotation_rate_tensor}\,.
\end{align}

The Euler equations governing momentum and internal energy conservation (Equations~2 and~3 in Paper~I) are correspondingly generalized to include the full stress tensor,

\begin{align}
\rho\frac{\upd v^{\alpha}}{\upd t}&=\frac{\partial \sigma^{\alpha\beta}}{\partial x^\beta}\label{eq:Momentum_conservation}\,,\\
\rho\frac{\upd u}{\upd t}&=\sigma^{\alpha\beta}\dot{\epsilon}^{\alpha\beta}\label{eq:Energy_conservation}\,.
\end{align}

\noindent Following the numerical approach described in Paper~I, these equations are discretized using SPH kernel summations. The resulting particle accelerations take the form

\begin{align}
&\frac{\upd v_i^\alpha}{\upd t}=\left.\frac{\upd v_i^\alpha}{\upd t}\right|_{H}+\left.\frac{\upd v_i^\alpha}{\upd t}\right|_{\Pi}+\left.\frac{\upd v_i^\alpha}{\upd t}\right|_{G}+\nonumber\\
&\sum\limits_{j}^Nm_j\left(\frac{S_i^{\alpha\beta}}{\Omega_i\rho_i^2}\frac{\partial W_{ij}(h_i)}{\partial x_i^\beta}+\frac{S_j^{\alpha\beta}}{\Omega_j\rho_j^2}\frac{\partial W_{ij}(h_j)}{\partial x_i^\beta}\right)\,,
\end{align}

\noindent while the corresponding equation for the internal energy becomes

\begin{equation}
\frac{\upd u_i}{\upd t}=\left.\frac{\upd u_i}{\upd t}\right|_H+\left.\frac{\upd u_i}{\upd t}\right|_\Pi-\frac{S_i^{\alpha\beta}}{\Omega_i\rho_i^2}\sum\limits_{j}^Nm_jv_{ij}^\alpha\frac{\partial W_{ij}(h_i)}{\partial x_i^\beta}\,,
\end{equation}

\noindent Here, the subscripts $H$, $\Pi$, and $G$ denote the hydrodynamic, artificial viscosity, and gravitational contributions, respectively.

The strain-rate and rotation-rate tensors require an estimate of the velocity gradient. In SPH, this is obtained via kernel summation as

\begin{align}
&\mathbb{F}_i^{\alpha\beta}=\frac{\partial v_i^\alpha}{\partial x_i^\beta}=\nonumber\\
&-\frac{1}{2}\sum\limits_{j}^Nm_jv_{ij}^\alpha\left(\frac{1}{\Omega_i\rho_i}\frac{\partial W_{ij}(h_i)}{\partial x_i^\beta}+\frac{1}{\Omega_j\rho_j}\frac{\partial W_{ij}(h_j)}{\partial x_i^\beta}\right)\,.\label{eq:Velocity_Gradient}
\end{align}

In this form, the discretization does not strictly conserve angular momentum. To restore first-order angular momentum conservation, we introduce a correction tensor $C_i^{\alpha\beta}$ \citep{bonetVariationalMomentumPreservation1999, chenImprovementTensileInstability1999, schaferCollisionsEqualsizedIce2007}, defined as

\begin{align}
&\left(C_i^{\alpha\beta}\right)^{-1}=\frac{\partial x_i^\alpha}{\partial x_i^\beta}=\nonumber\\
&-\frac{1}{2}\sum\limits_{j}^Nm_jx_{ij}^\alpha\left(\frac{1}{\Omega_i\rho_i}\frac{\partial W_{ij}(h_i)}{\partial x_i^{\beta}}+\frac{1}{\Omega_j\rho_j}\frac{\partial W_{ij}(h_j)}{\partial x_i^{\beta}}\right)\,.\label{eq:Correction_Tensor}
\end{align}

Multiplication of the raw velocity gradient tensor by this correction tensor yields the corrected gradient

\begin{equation}\label{eq:Corrected_Gradients}
\mathbb{G}_i^{\alpha\beta} = \mathbb{F}_i^{\alpha\gamma}C_i^{\gamma\beta}\,,
\end{equation}

\noindent which is subsequently used to compute the rotation-rate and strain-rate tensors according to

\begin{eqnarray}
R^{\alpha\beta}&=&\frac{1}{2}\left(\mathbb{G}_i^{\alpha\beta}-\mathbb{G}_i^{\beta\alpha}\right)\,,\\
\dot{\epsilon}^{\alpha\beta}&=&\frac{1}{2}\left(\mathbb{G}_i^{\alpha\beta}+\mathbb{G}_i^{\beta\alpha}\right)\,.
\end{eqnarray}

We would like to point out that the expression used to compute the gradients in Equations~\eqref{eq:Velocity_Gradient} and~\eqref{eq:Correction_Tensor} performs better at material interfaces compared to the non-symmetric form that is also available in \texttt{pkdgrav3}, but may introduce larger zeroth order errors in the gradient estimate.

Finally, by exploiting the symmetry and tracelessness of the deviatoric stress tensor, Equation~\eqref{eq:Deviatoric_Stress_derivative} can be written explicitly in component form for numerical integration as

\begin{eqnarray}
\dot{S}_i^{xx} &=& S_i^{xy}\left(\mathbb{G}_i^{yx}-\mathbb{G}_i^{xy}\right) + S_i^{xz}\left(\mathbb{G}_i^{zx} - \mathbb{G}_i^{xz}\right) \nonumber\\
&-& \frac{2}{3}\Gamma_i\left(\mathbb{G}_i^{yy} + \mathbb{G}_i^{zz} - 2\mathbb{G}_i^{xx} \right)\,,\\
\dot{S}_i^{yy} &=& S_i^{xy}\left(\mathbb{G}_i^{xy} - \mathbb{G}_i^{yx}\right) + S_i^{yz}\left(\mathbb{G}_i^{zy} - \mathbb{G}_i^{yz}\right) \nonumber\\
&-& \frac{2}{3}\Gamma_i\left(\mathbb{G}_i^{xx} + \mathbb{G}_i^{zz} - 2\mathbb{G}_i^{yy}\right)\,,\\
\dot{S}_i^{xy} &=& \Gamma_i\left(\mathbb{G}_i^{xy} + \mathbb{G}_i^{yx}\right)\nonumber\\
&+& \frac{1}{2}\left[S_i^{xx}\left(\mathbb{G}_i^{xy} - \mathbb{G}_i^{yx}\right) + S_i^{yy}\left(\mathbb{G}_i^{yx}-\mathbb{G}_i^{xy}\right) \right.\nonumber\\
&+&\left. S_i^{yz}\left(\mathbb{G}_i^{zx} - \mathbb{G}_i^{xz}\right) + S_i^{xz}\left(\mathbb{G}_i^{zy}-\mathbb{G}_i^{yz}\right)\right]\,,\\
\dot{S}_i^{xz} &=& \Gamma_i\left(\mathbb{G}_i^{xz} + \mathbb{G}_i^{zx}\right)\nonumber\\
&+& \frac{1}{2}\left[\left(S_i^{xx} + S_i^{yy}\right)\left(\mathbb{G}_i^{xz} - \mathbb{G}_i^{zx}\right) + S_i^{xx}\left(\mathbb{G}_i^{xz} - \mathbb{G}_i^{zx}\right) \right.\nonumber\\
&+& \left. S_i^{yz}\left(\mathbb{G}_i^{yx} - \mathbb{G}_i^{xy}\right) + S_i^{xy}\left(\mathbb{G}_i^{yz} - \mathbb{G}_i^{zy}\right)\right]\,,\\
\dot{S}_i^{yz} &=& \Gamma_i\left(\mathbb{G}_i^{yz} + \mathbb{G}_i^{zy}\right)\nonumber\\
&+&\frac{1}{2}\left[\left(S_i^{xx} + S_i^{yy}\right)(\mathbb{G}_i^{yz} - \mathbb{G}_i^{zy}) + S_i^{xz}\left(\mathbb{G}_i^{xy} - \mathbb{G}_i^{yx}\right) \right.\nonumber\\
&+& \left. S_i^{xy}\left(\mathbb{G}_i^{xz} - \mathbb{G}_i^{zx}\right) + S_i^{yy}\left(\mathbb{G}_i^{yz} - \mathbb{G}_i^{zy}\right)\right]\,.
\end{eqnarray}

The shear modulus $\Gamma$ is provided by the equation of state library \texttt{EOSlib} \citep{meierEOSResolutionConspiracy2021,meierEOSlib2021}. Assuming a Poisson ratio of $\nu=0.25$, the shear modulus can be derived from the bulk modulus $K = \rho c_s^2$ as

\begin{equation}
\Gamma = \frac{3(1-2\nu)}{2(1+\nu)}K\,.
\end{equation}

\noindent The choice of $\nu = 0.25$ is representative of many geological materials, including silicate rocks and ice, for which Poisson ratios typically lie in the range $0.2 \lesssim \nu \lesssim 0.3$ under relevant pressure and temperature conditions. This value therefore provides a reasonable default approximation. Alternatively, the shear modulus can be specified explicitly as a constant at runtime, allowing distinct values to be assigned to different materials within a given simulation.

\subsection{Plasticity}
The model described above represents a purely elastic material that would recover its original shape once all stresses are removed. Real materials, however, exhibit plastic behavior above a threshold stress, beyond which deformations become permanent. To capture this behavior, \texttt{pkdgrav3} implements a plasticity model \citep{benzImpactSimulationsFracture1994,collinsModelingDamageDeformation2004,jutziModelingAsteroidCollisions2015}. In this framework, stresses exceeding the yield strength $Y$ cause the deviatoric stress tensor to be reduced by rescaling it according to

\begin{equation}
S^{\alpha\beta} \rightarrow fS^{\alpha\beta}\,.\label{eq:Deviatoric_stress_rescaling}
\end{equation}

\noindent Following the formulation of \citet{collinsModelingDamageDeformation2004}, the scaling factor $f$ is given by

\begin{equation}
f=\min\left(\frac{Y}{\sqrt{J_2}},1\right)\,,
\end{equation}

\noindent where $J_2$ is the second invariant of the deviatoric stress tensor,

\begin{equation}
J_2=\frac{1}{2}S^{\alpha\beta}S^{\alpha\beta}\,.
\end{equation}

While the von Mises criterion with a constant yield threshold is appropriate for ductile metals, geological materials such as rock and ice exhibit more complex behavior. At sufficiently large strains, these materials undergo brittle failure and fracture, reducing their ability to support shear stress. This behavior is commonly represented through a reduction in the effective yield strength. In addition, the high overburden pressures present in planetary interiors significantly increase material strength. Both effects are incorporated using the pressure-dependent yield strength model of \citet{collinsModelingDamageDeformation2004}. For intact (undamaged) material, the yield strength $Y_i$ is given by a smooth fit to experimental data:

\begin{equation}
Y_i=Y_0+\frac{\mu_iP}{1+\frac{\mu_iP}{Y_m-Y_0}}\,,
\end{equation}

\noindent where $\mu_i$ is the coefficient of internal friction for an intact material, $Y_0$ is the zero-pressure yield strength (cohesion), and $Y_m$ is the yield strength at infinite pressure, known as the \emph{von Mises plastic limit}. In the current implementation, pressures below zero are clipped to zero to avoid numerical problems \citep[e.g.][]{dehnenImprovingConvergenceSmoothed2012} and $Y_0$ is set to zero. For fully damaged (fractured, granular) material, a Coulomb dry-friction law is assumed:

\begin{equation}
Y_d=\mu_dP\,,
\end{equation}

\noindent where $\mu_d$ is the coefficient of internal friction for damaged material.

The current implementation in \texttt{pkdgrav3} does not include a damage model and therefore the yield strength used in the simulations is taken to be the minimum of the intact and damaged values,

\begin{equation}
Y = \min(Y_i,Y_d)\,,
\end{equation}

\noindent effectively modeling the material as fully damaged, following \citet{collinsModelingDamageDeformation2004} which is a good approximation for intermediate mass bodies \citep{emsenhuberSPHCalculationsMarsscale2018,ballantyneInvestigatingFeasibilityImpactinduced2023,ballantyneSputnikPlanitiaImpactor2024,dentonCaptureAncientCharon2025}.

Pressure is not the only factor influencing the yield strength of solid materials. As a material approaches its melting temperature, its shear strength decreases, ultimately vanishing in the fluid limit. This thermal softening is approximated using the formulation of \citet{collinsModelingDamageDeformation2004}, following \citet{ohnakaShearFailureStrength1995}:

\begin{equation}
Y\rightarrow Y\tanh\left(\xi\left(\frac{T_{melt}}{T}-1\right)\right)\,,
\end{equation}

\noindent where $T$ is the temperature, $T_{melt}$ is the melting temperature, and $\xi$ is a material-dependent thermal softening parameter.

In practice, the yield strength $Y$ is computed by \texttt{EOSlib} as a function of density $\rho$ and internal energy $u$ from which temperature $T$ and pressure $P$ are calculated. The material parameters entering the intact, damaged, and thermal softening formulations, $Y_0$, $Y_m$, $\mu_i$, $\mu_d$, and $\xi$, are specified as constant values at runtime for each material. The (pressure-dependent) melting temperature $T_{melt}$ is provided as a tabulated function of density $\rho$, either derived from the phase information of the (M)-ANEOS equation of state or obtained from analytic expressions derived from experimental or simulation data. In the latter case, melting curves typically given as $T(P)$ are converted to $T(\rho)$ using the equation of state. The resulting yield strength is then used in \texttt{pkdgrav3} to rescale the deviatoric stress according to Equation~\eqref{eq:Deviatoric_stress_rescaling}.

\subsection{Time integration}
The deviatoric stress tensor is integrated in time using the same predictor-corrector scheme that is employed for the internal energy (see Paper~I). The timestep size is determined using the hierarchical individual timestep scheme described in Paper~I. In particular, the timestep is limited by both an acceleration criterion,

\begin{equation}
\Delta t_{i,a}=\eta_a\sqrt{\frac{\epsilon}{\left\vert a_i\right\vert}}\,,
\end{equation}

\noindent and the Courant-Friedrichs-Lewy (CFL) condition,

\begin{equation}\label{eq:courant_condition}
\Delta t_{i,CFL}=\eta_C\min_j\frac{h_i}{(1+0.6\alpha)c_i+0.6\beta\left\vert\min(0,\mu_{ij})\right\vert}\,,
\end{equation}

\noindent where $\eta_a$ and $\eta_C$ are timestep parameters, $\epsilon$ is the gravitational softening length or SPH kernel size, $h_i$ is the smoothing length, $c_i$ is the sound speed, and $\alpha$ and $\beta$ are the artificial viscosity parameters. The timestep used for integration is the minimum of these criteria.

In the prediction step, the tensor is predicted from time $i-\frac{1}{2}$ to time $i$ using the time derivatives evaluated at $i-1$,

\begin{equation}
S_{i,pred}^{\alpha\beta}=S_{i-\frac{1}{2}}^{\alpha\beta}+\frac{1}{2}\Delta t\dot{S}_{i-1}^{\alpha\beta}\,.\label{eq:Predict_S}
\end{equation}

\noindent The predicted deviatoric stress tensor is then rescaled according to Equation~\eqref{eq:Deviatoric_stress_rescaling}, after which the updated time derivatives $\dot{S}_i^{\alpha\beta}$ are computed from the predicted values. In the closing kick, the tensor is advanced from time $i-\frac{1}{2}$ to time $i$ using these new derivatives,

\begin{equation}
S_{i}^{\alpha\beta}=S_{i-\frac{1}{2}}^{\alpha\beta}+\frac{1}{2}\Delta t\dot{S}_i^{\alpha\beta}\,,
\end{equation}

\noindent and the rescaling given by Equation~\eqref{eq:Deviatoric_stress_rescaling} is applied again. Finally, the opening kick advances the tensor to time $i+\frac{1}{2}$,

\begin{equation}
S_{i+\frac{1}{2}}^{\alpha\beta}=S_{i}^{\alpha\beta}+\frac{1}{2}\Delta t\dot{S}_i^{\alpha\beta}\,.
\end{equation}

The code also tracks the second invariant of the strain-rate tensor by integrating the scalar quantity constructed from the strain-rate tensor defined in Equation~\eqref{eq:Strain_rate_tensor},
\begin{equation}
\frac{1}{2}\dot{\epsilon}^{\alpha\beta}\dot{\epsilon}^{\alpha\beta}\,,
\end{equation}
which provides a rotationally invariant measure of the accumulated strain rate.

\section{Tests with strength}\label{sec:Tests_with_strength}
In this section, we validate the shear strength implementation by comparing simulation results with laboratory experiments. We then present two additional simulation scenarios that illustrate the qualitative and quantitative differences between simulations performed with and without the shear strength model, highlighting its impact on the resulting material response and dynamical evolution. Finally, we assess the conservation properties of the method by examining angular momentum conservation in a representative impact scenario. All simulations were carried out using the Wendland C2 kernel with 100 neighbors \citep{dehnenImprovingConvergenceSmoothed2012} in a unit system defined by $[L] = \SI{1}{\Rearth}$, $[V] = \SI{1}{\kilo\meter\per\second}$, and $[G] = 1$. For the equation of state, we adopt M-ANEOS quartz \citep{thompsonImprovementsCHARTRadiationhydrodynamic1974, meloshHydrocodeEquationState2007, thompsonMANEOS2019, meierANEOSmaterial2021}, with strength parameters $\Gamma = \SI{72}{\giga\pascal}$, $Y_0 = \SI{0}{\giga\pascal}$, $Y_m = \SI{3.5}{\giga\pascal}$, $\mu_i = 2.0$, $\mu_d = 0.8$ (unless stated otherwise), and $\xi = 1.2$. For the melt curve we use the data for $SiO_2$ obtained from \citet{gonzalez-cataldoMeltingCurveSiO22016}.

\subsection{Cliff collapse}\label{sec:Cliff_collapse}

\begin{figure}[ht!]
\centering
\includegraphics[width=\linewidth]{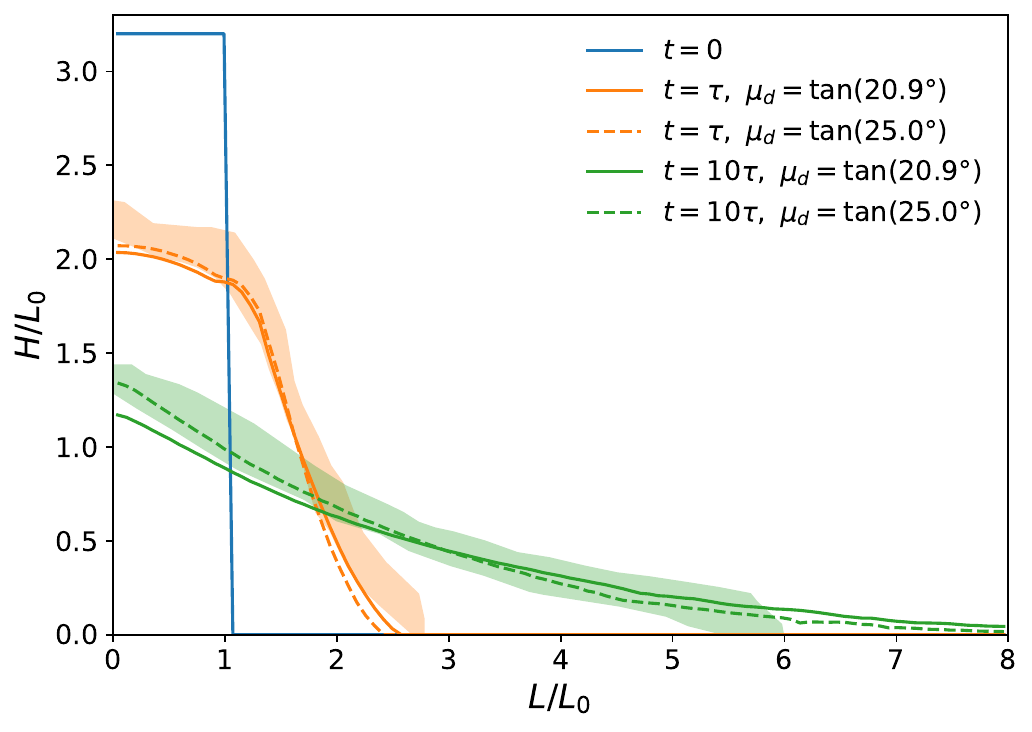}
\caption{Comparison between SPH simulations and laboratory experiments for the collapse of a cliff of granular material. The shaded regions (orange and green) show the experimental profiles of \citet{lajeunesseGranularSlumpingHorizontal2005} at $t = \tau$ and $t = \infty$, respectively, scaled by the initial length $L_0$ and characteristic time $\tau = \sqrt{H_0/g}$. Solid lines show the SPH simulation results for $\mu_d = \tan{\SI{20.9}{\degree}}$, and dashed lines show the results for $\mu_d = \tan{\SI{25}{\degree}}$. The curves are constructed by binning the particles along the $x$-axis into 101 bins spanning the interval $[0, 8L_0]$ and taking the maximum $y$-value within each bin. Both friction values reproduce the experimental behavior, with the dashed lines ($\mu_d = \tan{\SI{25}{\degree}}$) providing an almost perfect match to the final run-out profile.}
\label{fig:cliff_collapse_profiles}
\end{figure}

Similar to \citet{jutziSPHCalculationsAsteroid2015}, we model the collapse of a cliff of granular material and validate our implementation of pressure dependent shear strength (i.e., friction) against the experiments of \citet{lajeunesseGranularSlumpingHorizontal2005}. We note that \citet{holsappleModelingGranularMaterial2013} performed a detailed study of the same problem using a continuum CTH code with Mohr-Coulomb and/or Drucker-Prager constitutive models.

In the cliff collapse setup (see \citet{holsappleModelingGranularMaterial2013} for a detailed description), granular material is initially confined to a rectangular region of height $H_0$ and length $L_0$, with a width assumed sufficiently large to be of negligible effect on the results. At $t=0$, one of the side walls is removed, allowing the material to flow downward under external gravity $\vec{g}$. The flow eventually comes to rest in a final run-out configuration with maximum height $H$ and length $L$, which depend primarily on the material properties, notably the internal angle of friction. Both experiments \citep{lajeunesseGranularSlumpingHorizontal2005} and scaling analyses \citep{holsappleModelingGranularMaterial2013} show that the final scaled profile depends only on the initial height-to-length ratio and the friction angle; the absolute values of gravity and cliff height influence only the characteristic time and length scales. This allows simulations to be performed using artificially large dimensions, mitigating timestep restrictions (see \cite{holsappleModelingGranularMaterial2013}). In the 3D calculation presented here, we set $H_0 = \SI{10}{\kilo\meter}$. The granular material is sampled with $512 \times 160$ BCC cells in the $H$ and $L$ directions (see Paper I), resulting in an aspect ratio of 3.2, consistent with the experiments of \citet{lajeunesseGranularSlumpingHorizontal2005}. In the perpendicular direction, 16 cells are used, confined by periodic boundaries. The surface at $y=0$ is represented by 10 layers of fixed particles. To prevent artifacts from particles sticking to the boundary at $x=0$, we instead implement a mirrored boundary by additionally sampling a mirrored copy of the material.

Figure~\ref{fig:cliff_collapse_profiles} compares the SPH simulation results with the experimental measurements of \citet{lajeunesseGranularSlumpingHorizontal2005}. In the experiments, glass beads of various sizes and initial height-to-length ratios were used. When scaled by the characteristic length $L_0$ and time $\tau = \sqrt{H_0/g}$, the resulting profiles are nearly indistinguishable. These measured curves are represented by the orange and green shaded regions in Figure~\ref{fig:cliff_collapse_profiles} at $t = \tau$ and $t = \infty$, respectively. 

Our simulations were performed using two different friction coefficients, $\mu_d$. The solid lines correspond to $\mu_d = \tan{\SI{20.9}{\degree}}$, the value derived for $\mu_s$ by \citet{jopConstitutiveLawDense2006}, and the dashed lines correspond to $\mu_d = \tan{\SI{25}{\degree}}$, following \citet{holsappleModelingGranularMaterial2013}. Both choices reproduce the experimental results accurately, with $\mu_d = \tan{\SI{25}{\degree}}$ providing an almost perfect match to the final run-out profile at $t = 10 \tau$, once all motion has ceased.

\subsection{Catastrophic disruption}\label{sec:Catastrophic_disruption}
For the evolution of small bodies and growing planets, erosive or disruptive collisions play a key role. A commonly used measure of the erosiveness of a collision is its comparison to the catastrophic disruption threshold $Q_{RD}^*$, defined as the specific impact energy required to unbind or remove \SI{50}{\percent} of the initial total mass \citep{benzCatastrophicDisruptionsRevisited1999,stewartVELOCITYDEPENDENTCATASTROPHICDISRUPTION2009,leinhardtCollisionsGravitydominatedBodies2012}.

The specific impact energy in the center-of-mass frame is given by

\begin{equation}
Q_R=\frac{1}{2}\frac{M_{tar}M_{imp} v^2}{(M_{tar}+M_{imp})^2}\,,\label{eq:Q_R}
\end{equation}

\noindent where $M_{tar}$ and $M_{imp}$ denote masses of the target and impactor, respectively, and $v$ is the impact velocity.

Although $Q_{R}$ is normalized by mass, $Q_{RD}^*$ still exhibits a strong dependence on the mass/size regime. It also depends on the mass ratio, the impact velocity and the material rheology, i.e. whether the material is modeled as a strengthless fluid or as a granular solid with finite strength. This rheological dependence itself varies with mass: for sufficiently large bodies, $Q_{RD}^*$ becomes effectively independent of material strength \citep{benzCatastrophicDisruptionsRevisited1999,jutziModelingAsteroidCollisions2015}. This behavior has long been used to justify neglecting material strength in simulations of giant impacts involving bodies significantly larger than what is sometimes called the potato radius $r_{potato}\simeq$ \SIrange{200}{300}{\kilo\meter} \citep{lineweaverPotatoRadiusLower2010}.

As Equation~\eqref{eq:Q_R} shows, for a given total mass $M_{tot}=M_{tar}+M_{imp}$, the specific impact energy depends on how the mass is partitioned between the two bodies as well as on the impact velocity. Introducing the mass ratio $\gamma=M_{imp}/M_{tar}$, Equation~\eqref{eq:Q_R} can be rewritten as

\begin{equation}
Q_R=\frac{1}{2}\frac{\gamma}{(\gamma+1)^2}v^2\,,
\end{equation}

\noindent where the explicit dependence on the total mass cancels out.

To determine $Q_{RD}^*$ for a given total mass, a series of simulations with varying $Q_R$ must be performed in order to identify the value that results in the loss of half the initial mass. Given the dependence of $Q_{R}$ on $\gamma$ and $v$, two complementary approaches can be used to vary $Q_R$. For low-mass bodies, $Q_{RD}^*$ is small and catastrophic disruption typically occurs for impactors much smaller than the target. In this regime, $Q_R$ is usually varied by changing $\gamma$ at a fixed impact velocity. For larger total masses, a second approach becomes more practical, in which the mass ratio is held fixed and $Q_R$ is varied by changing the impact velocity.

For each total mass, we determined $Q_{RD}^*$ using a bisection search in $Q_R$. The particle representation of the bodies was created from a body-centered cubic (BCC) lattice at the reference density of quartz, $\rho_0=\SI{2.65}{\gram\per\centi\meter\tothe3}$, with a homogeneous temperature of \SI{300}{\kelvin}. From this lattice, spherical target and impactor bodies were extracted such that either the total number of particles in a simulation was $N=10^6$ in the high-mass regime, where the impact velocity $v$ is varied, or the target contained $N=10^6$ particles in the low-mass regime, where the mass ratio $\gamma$ is varied. For very small values of $\gamma$, the impactor was represented by a minimum of $N=341$ particles, with the particle mass adjusted to achieve the desired mass ratio. The larger bodies were subsequently evolved in isolation using a ramped velocity damper (see Paper~I) to reach a relaxed equilibrium state. The impact simulations were performed as head-on collisions and evolved for sufficiently long durations to allow the collision outcome to fully develop and for bound and unbound material to be robustly distinguished. For each output snapshot, the bound mass was determined using \texttt{SKID} \citep{n-bodyshopSKIDFindingGravitationally2011} by iteratively removing unbound particles until convergence. Detailed simulation parameters, including the target and impactor masses, impact velocities, and the procedure used to determine $Q_{RD}^*$ from the simulation results, are provided in Appendix~\ref{sec:appendix:QRDstar_tables}.

\begin{figure}[ht!]
\centering
\includegraphics[width=\linewidth]{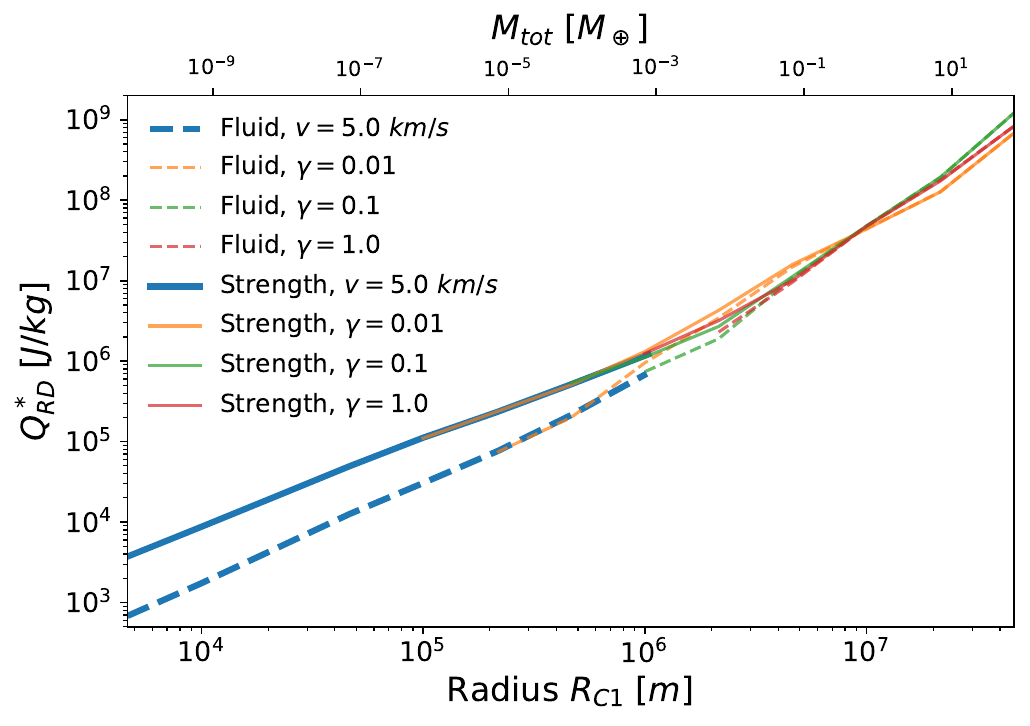}
\caption{Catastrophic disruption threshold $Q_{RD}^*$ as a function of total body mass, parameterized by the equivalent radius $R_{C1}$ of a sphere with density \SI{1}{\gram\per\centi\meter\tothe3}. Results are shown for simulations with material strength (solid lines) and for a strengthless fluid (dashed lines). Results for constant impact velocity are show with the blue lines, while those for constant mass ratios $\gamma$ are shown in orange, green and red. Material strength significantly increases $Q_{RD}^*$ at low masses, while the strength and fluid results converge at larger masses.}
\label{fig:QRDstar}
\end{figure}

\begin{figure}[ht!]
\centering
\includegraphics[width=\linewidth]{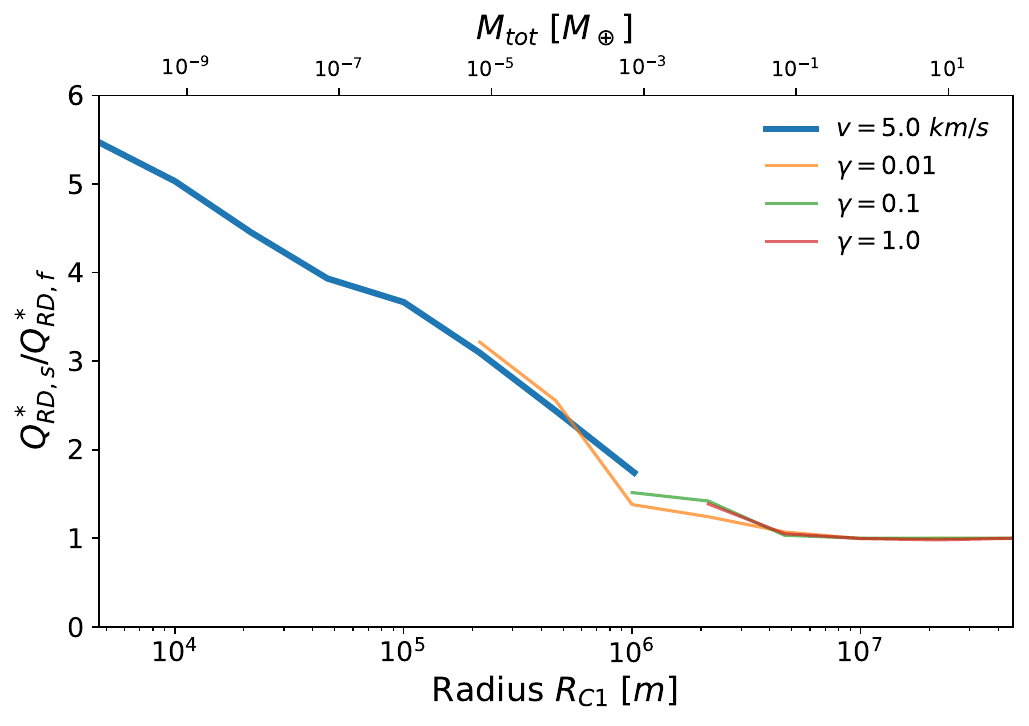}
\caption{Ratio of the catastrophic disruption threshold $Q_{RD}^*$ obtained from simulations with material strength to that obtained from fluid simulations, shown as a function of total mass parameterized by $R_{C1}$. The ratio decreases monotonically with increasing mass and reaches unity at $R_{C1}=\SI{1e7}{\meter}$, corresponding to a mass of approximately \SI{0.7}{\Mearth}. This indicates that the influence of material strength on catastrophic disruption diminishes with increasing body size and becomes negligible only at sizes exceeding a few thousand kilometers.}
\label{fig:QRDstar_ratio}
\end{figure}

Figure~\ref{fig:QRDstar} presents the resulting values of $Q_{RD}^*$ over a wide range of masses, parameterized by the radius of a sphere with density \SI{1}{\gram\per\centi\meter\tothe3} (known as $R_{C1}$), for simulations with and without material strength. In the low-mass regime, $Q_R$ is varied by changing $\gamma$ at a constant impact velocity of $v=\SI{5}{\kilo\meter\per\second}$, up to a maximum impactor mass of \SI{10}{\percent} of the target mass. In the high-mass regime ($R_{C1}>\SI{1000}{\kilo\meter}$, $M_{tot}>\SI{7e-4}{\Mearth}$), the alternative approach is used: $\gamma$ is fixed at three different values and $v$ is varied, starting at $v\geq\SI{3}{\kilo\meter\per\second}$, corresponding to the sound speed of the quartz material used in the simulations.

Significant differences between the strength and fluid models are evident in the low-mass regime, where material strength substantially increases $Q_{RD}^*$. With increasing mass, these differences steadily diminish, and the two models converge and fully coincide at large masses. In addition, the two approaches used to vary $Q_R$ show a general agreement in the region where they overlap, where impact velocities are \SIrange{3}{5}{\kilo\meter\per\second}.

Figure~\ref{fig:QRDstar_ratio} shows the ratio of $Q_{RD}^*$ obtained with material strength to that obtained using the fluid model as a function of total mass. This ratio decreases monotonically with increasing mass and reaches unity at $R_{C1}=\SI{1e7}{\meter}$ corresponding to a mass of approximately \SI{0.7}{\Mearth}. Notably, this transition occurs at sizes exceeding a few thousand kilometers, indicating that the influence of material strength persists to significantly larger bodies than is often assumed, consistent with previous work demonstrating the importance of strength in impacts at Mars scales \citep{golabekCouplingSPHThermochemical2018,emsenhuberSPHCalculationsMarsscale2018,ballantyneInvestigatingFeasibilityImpactinduced2023}.

\begin{figure}[ht!]
\centering
\includegraphics[width=\linewidth]{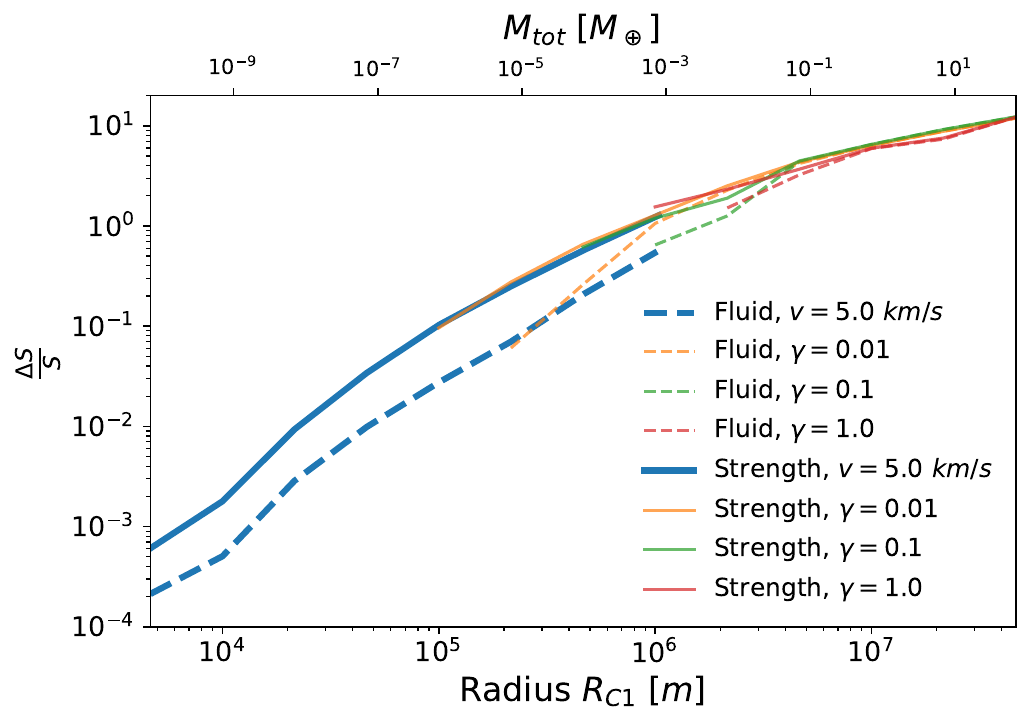}
\caption{Relative increase in entropy of the bound material at the catastrophic disruption threshold $Q_{RD}^*$ as a function of total mass, comparing the simulations with material strength to those without. At low masses, the inclusion of material strength leads to substantially higher entropy production, while at larger masses the entropy increase converges toward the result without strength.}
\label{fig:QRDstar_deltaS}
\end{figure}

\begin{figure}[ht!]
\centering
\includegraphics[width=\linewidth]{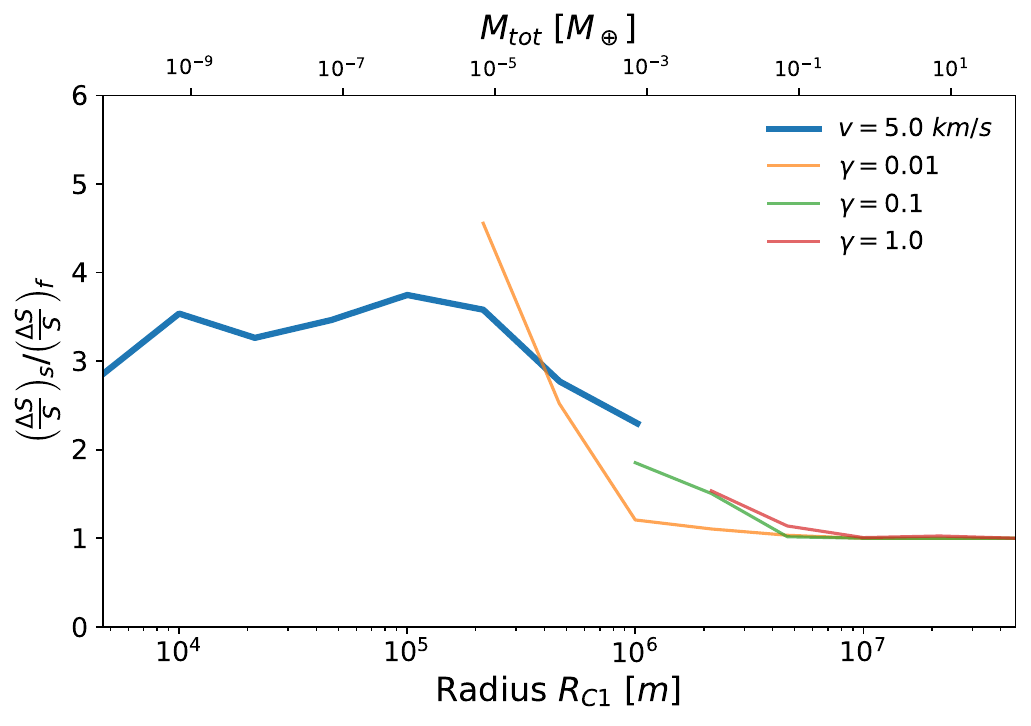}
\caption{Ratio of the relative entropy increase of the bound material at $Q_{RD}^*$ obtained with material strength to that obtained using the fluid model, shown as a function of total mass. The ratio approaches unity at large masses, indicating convergence between the two rheological treatments, while substantial deviations at low masses highlight the importance of material strength for entropy generation in smaller bodies.}
\label{fig:QRDstar_deltaS_ratio}
\end{figure}

Figures~\ref{fig:QRDstar_deltaS} and~\ref{fig:QRDstar_deltaS_ratio} show the relative increase in entropy of the bound material at the catastrophic disruption threshold $Q_{RD}^*$ for simulations with and without material strength, and the ratio between these values, respectively. The entropy increase exhibits behavior broadly similar to that of $Q_{RD}^*$, with large differences between strength and fluid models at low masses and convergence toward a common value at higher masses. Even though all simulations use the exact entropy conserving scheme, which substantially reduces noise in the entropy, the entropy increase shows significantly more scatter than $Q_{RD}^*$, reflecting its greater sensitivity to local shock structure, numerical noise, artificial viscosity treatment and small-scale variations in impact geometry and material response.

It is important to emphasize that the convergence of $Q_{RD}^*$ between strength and fluid models at large planetary masses does not imply that material rheology is irrelevant. For sufficiently massive bodies, the interior pressure due to self-gravity exceeds the material yield strength, making gravity the dominant force and allowing the body to be approximately described by a fluid rheology for the purposes of determining $Q_{RD}^*$. In this regime, the equation of state becomes the primary factor controlling impact response, particularly for shock compression, melting, and vaporization. Nonetheless, material strength still influences the propagation and decay of shock waves, as well as localized features such as basin formation, because the speed of sound and shock characteristics differ between solid and liquid material. At intermediate scales, such as Mars-, Mercury-, and Moon-sized bodies, both self-gravity and material strength play comparable roles, and accurate modeling requires accounting for their combined effects \citep[e.g.,][]{emsenhuberSPHCalculationsMarsscale2018, ballantyneInvestigatingFeasibilityImpactinduced2023, ballantyneSputnikPlanitiaImpactor2024, dentonCaptureAncientCharon2025}.

\subsection{Influence of material strength on remnant shape}
\begin{figure*}[ht!]
\centering
\includegraphics[width=\linewidth]{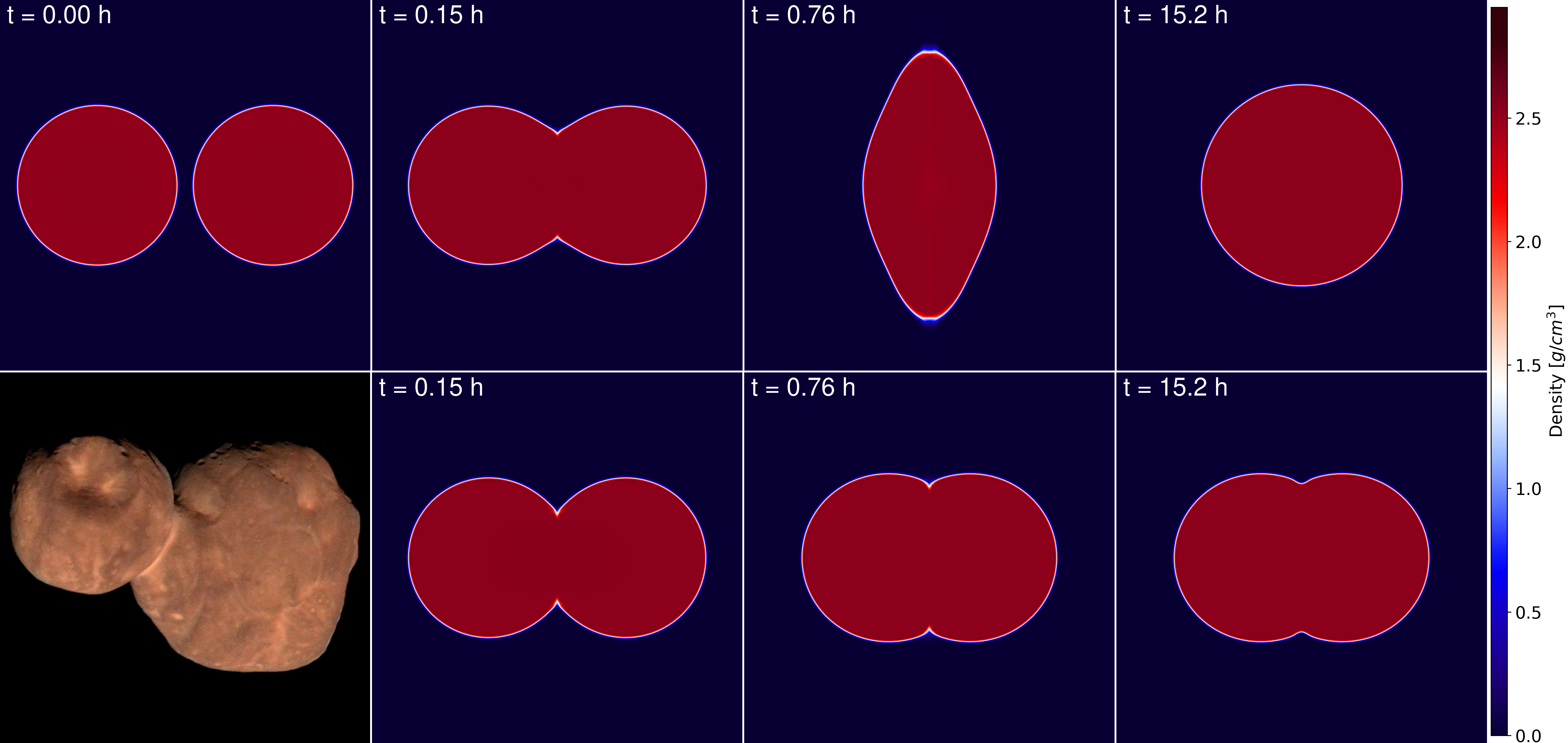}
\caption{Time series of density slices through a low-velocity collision ($v = \SI{100}{\meter\per\second}$) for bodies from the $Q_{RD}^*$ runs with $\gamma=1.0$ and $R_{C1} = \SI{2.15e5}{\meter}$ (each body having a radius of \SI{123.5}{\kilo\meter} and a mass of \SI{3.5e-6}{\Mearth}) at a resolution of $10^7$ particles, comparing simulations with and without material strength. The density is calculated by distributing the mass of each particle to each pixel according to the intersection between the pixel area and the particle's kernel using the SPH kernel function. In the simulation without strength (top row), the remnant rapidly relaxes toward a nearly spherical shape, exhibiting fluid-like behavior. In the simulation with strength (bottom row, for the strength parameters used see Section~\ref{sec:Tests_with_strength}), the remnant preserves a bi-lobed structure, with the two lobes remaining in contact at the point where they were initially compressed together. To illustrate such a bi-lobed object, the bottom-left panel shows an image of the Kuiper Belt object (486958) Arrokoth obtained by the New Horizons mission (Image credit: NASA, JHU APL, Southwest Research Institute, ESA). We emphasize, however, that Arrokoth is significantly smaller (diameter $\sim\SI{31.7}{\kilo\meter}$) and likely less dense than the bodies considered here. Accurately modeling such an object would therefore require the inclusion of additional physical effects, such as porosity, as explored, e.g., by \citet{marohnicConstrainingFinalMerger2021}.}
\label{fig:Composite_Figure_stick}
\end{figure*}

The shapes of planetary bodies are strongly influenced by their material strength properties. At sufficiently low masses, where the gravitational overburden is smaller than or comparable to the material yield strength, bodies can maintain highly irregular geometries. With increasing mass, self-gravity progressively dominates over strength, leading to more rounded shapes. Observations of asteroids and small bodies in the solar system reveal a remarkable diversity of morphologies, reflecting this interplay between gravity and material strength \citep[e.g.,][]{jutziShapeStructureCometary2015,leleuPeculiarShapesSaturns2018,raducanMultipleMoonletMergers2025}.

In our simulations, we confirm this sensitivity of post-impact shape to material strength. Figure~\ref{fig:Composite_Figure_stick} shows a time series of a low-velocity collision ($v=\SI{100}{\meter\per\second}$) for bodies taken from the $Q_{RD}^*$ runs with $\gamma=1$ and $R_{C1} = \SI{2.15e5}{\meter}$ (each body having a radius of \SI{123.5}{\kilo\meter} and a mass of \SI{3.5e-6}{\Mearth}), comparing simulations with and without shear strength.

In the simulation employing a fluid rheology (i.e., without shear strength), the remnant rapidly relaxes toward a nearly spherical configuration, as material flows to minimize gravitational potential energy. By contrast, in the simulation with shear strength, a bi-lobed structure is preserved: the two lobes remain in contact at the location where they were initially compressed together. The resistance to shear deformation and internal rearrangement inhibits large-scale relaxation toward spherical symmetry. Shapes like this have been observed in nature, for example, in cometary nuclei, small asteroids and moons, and reproduced in other SPH simulations \citep[e.g.,][]{jutziShapeStructureCometary2015,leleuPeculiarShapesSaturns2018,jutziShapeStructureSmall2019,wimarssonDiverseShapesBinary2025,raducanMultipleMoonletMergers2025}.

The differences in morphology are not only visually striking but also physically significant. In addition to shear strength, other material properties such as porosity, cohesion, and damage evolution are important for a fully realistic representation of asteroid and cometary interiors. Even in isolation, however, shear strength has a profound impact: it governs stress transmission and resists internal deformation, preserving irregular shapes that would otherwise relax toward spherical symmetry under self-gravity. This effect is particularly relevant for understanding contact binaries, elongated small bodies, and other irregular objects in the solar system, demonstrating that strength-dominated rheology fundamentally alters post-impact morphology relative to purely fluid models.

\subsection{Angular momentum conservation}
\begin{figure}[ht!]
\centering
\includegraphics[width=\linewidth]{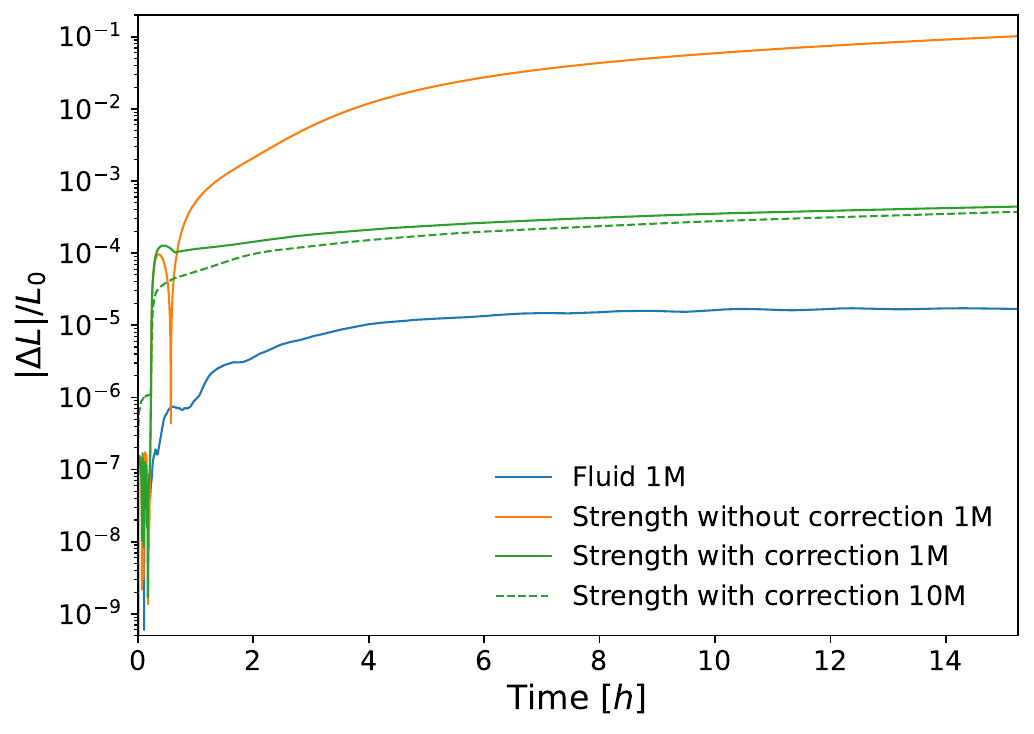}
\caption{Angular momentum conservation error, $\vert\Delta L\vert / L_0$, during an oblique impact with an impact angle of \SI{45}{\degree} between the same bodies as in Figure~\ref{fig:Composite_Figure_stick}. In the absence of material strength, the error remains at the level of \SI{e-5}{}. Including strength without the correction tensor (see Equation~\eqref{eq:Corrected_Gradients}) results in errors exceeding \SI{1}{\percent} after one and \SI{10}{\percent} after three rotations. The inclusion of the correction tensor substantially improves angular momentum conservation, although the error remains larger than in the strength-free case. Higher resolution leads to a slight additional improvement.}
\label{fig:AM_conservation}
\end{figure}

As discussed in Section~\ref{sec:Shear_strength}, a correction tensor is applied to the velocity gradient tensor (see Equation~\eqref{eq:Corrected_Gradients}) to ensure angular momentum conservation. To assess the effectiveness of this correction, we simulate an oblique impact between the same bodies as in Figure~\ref{fig:Composite_Figure_stick}, using an impact angle of \SI{45}{\degree} and an impact velocity of $v=\SI{100}{\meter\per\second}$. We then track the evolution of the total angular momentum throughout the simulation.

When shear strength is included, the impact produces a bi-lobed remnant with a rotation period of approximately \SI{5}{\hour}, corresponding to about three full rotations over the course of the simulation. In contrast, as expected, neglecting strength yields a nearly spherical remnant, consistent with Figure~\ref{fig:Composite_Figure_stick}. To isolate the effect of the correction tensor, we compare simulations performed without strength, with strength but without the correction tensor (by setting $C_i^{\alpha\beta} = \mathds{1}$), and with both strength and the correction tensor.

Figure~\ref{fig:AM_conservation} shows the resulting evolution of the relative angular momentum error, $\vert\Delta L\vert / L_0$, where

\begin{equation}
\vec{L} = \sum_i m_i \vec{r}_i\times \vec{v}_i\,,
\end{equation}

\noindent relative to the origin of the global coordinate system, which in this case coincides with the center of mass. Without strength, the angular momentum error remains at the level of \SI{e-5}{}. Including strength but omitting the correction tensor leads to a severe violation of angular momentum conservation, with errors already exceeding \SI{1}{\percent} after one and \SI{10}{\percent} after three rotations. Applying the correction tensor significantly improves conservation, reducing the error to below \SI{e-3}{}, although it remains worse than in the strength-free case, as expected. Increasing the numerical resolution yields a modest further improvement.

\section{Performance}\label{sec:Performance}

\begin{figure}[ht!]
\centering
\includegraphics[width=\linewidth]{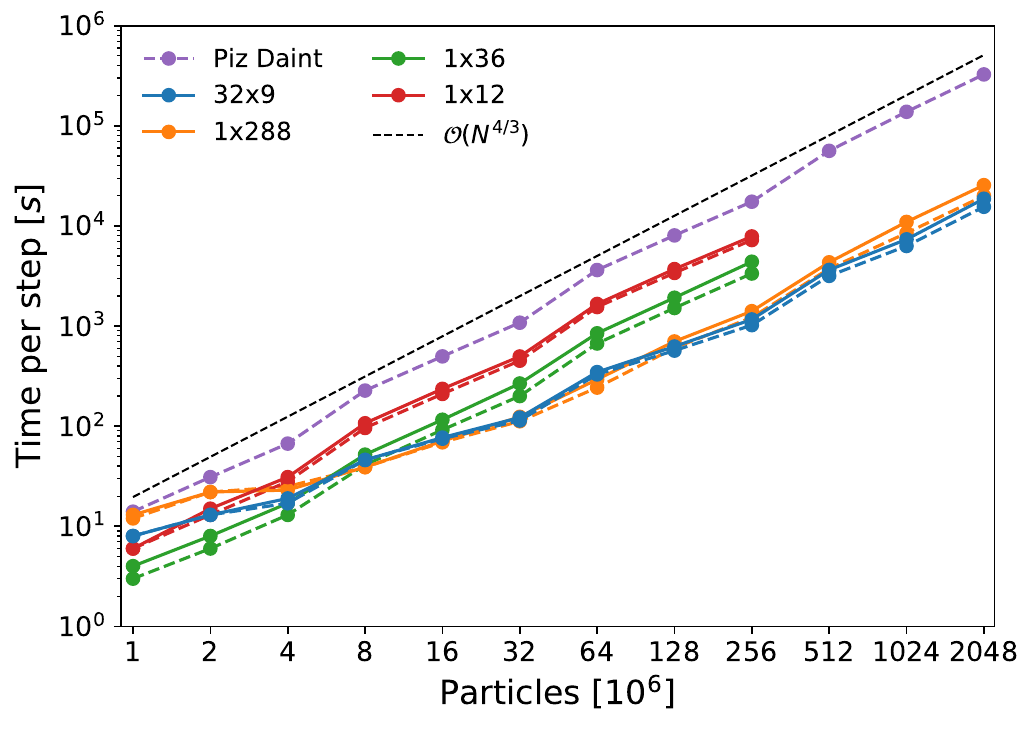}
\caption{Scaling of the time per step for the benchmark problem used in Paper~I, comparing results obtained on Piz Daint and on the ALPS system. Results with the shear strength model enabled are shown as solid lines, and results without the model are shown as dashed lines. Shown are ALPS runs using 12 cores ($1\times12$), 36 cores ($1\times36$), 288 cores (full-node utilization in a single MPI rank, $1\times288$), and a hybrid configuration with 32 MPI ranks and 9 worker threads per rank ($32\times9$). Configurations using a single compute module ($1\times12$ and $1\times36$) are omitted at resolutions exceeding the module memory. Piz Daint reference results are shown in purple, while the 12-core ALPS results are shown in red, illustrating the substantially improved per-core performance of the newer architecture. For the 12- and 36-core configurations, the time per step follows the expected $\mathcal{O}\bigl(N^{4/3}\bigr)$ scaling, whereas full-node configurations deviate at low resolutions due to domain decomposition and communication overheads. Enabling the shear strength model increases the computational cost across all configurations.}
\label{fig:scaling_comparison}
\end{figure}

\begin{figure}[ht!]
\centering
\includegraphics[width=\linewidth]{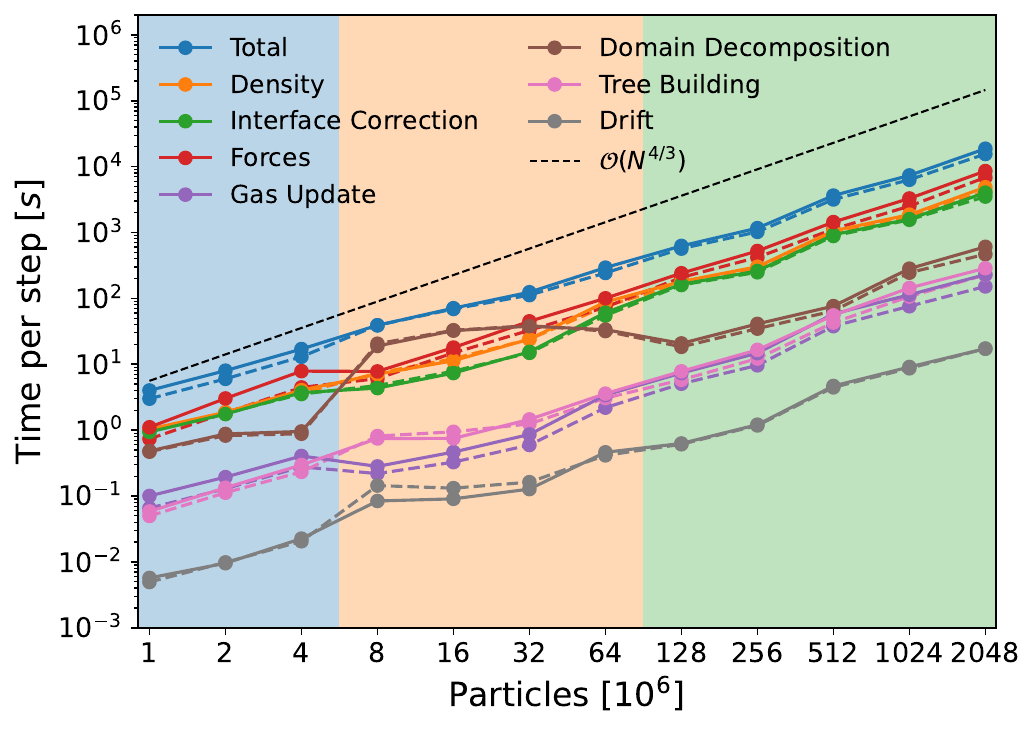}
\caption{Time per step for the best-performing configuration at each resolution on ALPS, with (solid lines) and without (dashed lines) the shear strength model enabled, including a breakdown of the contributions from the individual computational operations. Three resolution regimes can be identified: below $8\times10^{6}$ particles, the 36-thread configuration yields the lowest time per step; between $8\times10^{6}$ and $6.4\times10^{7}$ particles, the single-rank 288-thread configuration is optimal; and above $6.4\times10^{7}$ particles, the hybrid configuration with 32 MPI ranks and 9 worker threads per rank achieves the best performance by more effectively saturating the four GPUs. Tree-walk operations (density, interface correction, and forces) dominate the total step time, while domain decomposition, tree building, and the gas update contribute less, and the drift is negligible. The contribution from domain decomposition strongly varies with configuration and becomes dominant at low resolutions for the 288-thread setup.}
\label{fig:scaling_best}
\end{figure}

\begin{figure*}[ht!]
\centering
\includegraphics[width=0.5\linewidth]{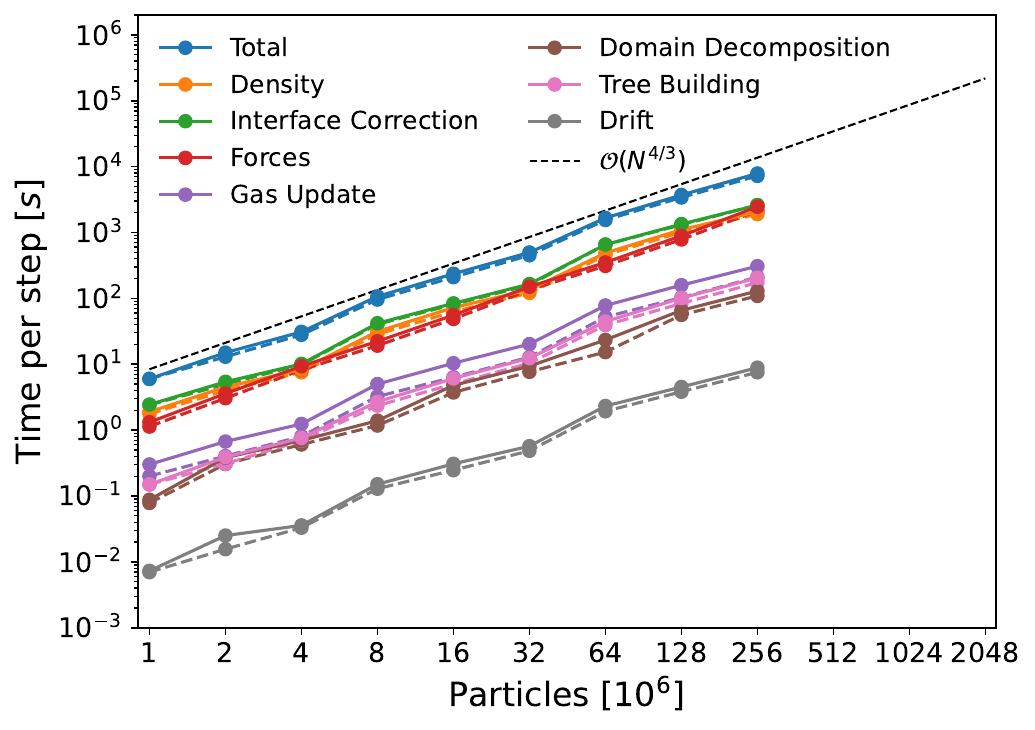}\includegraphics[width=0.5\linewidth]{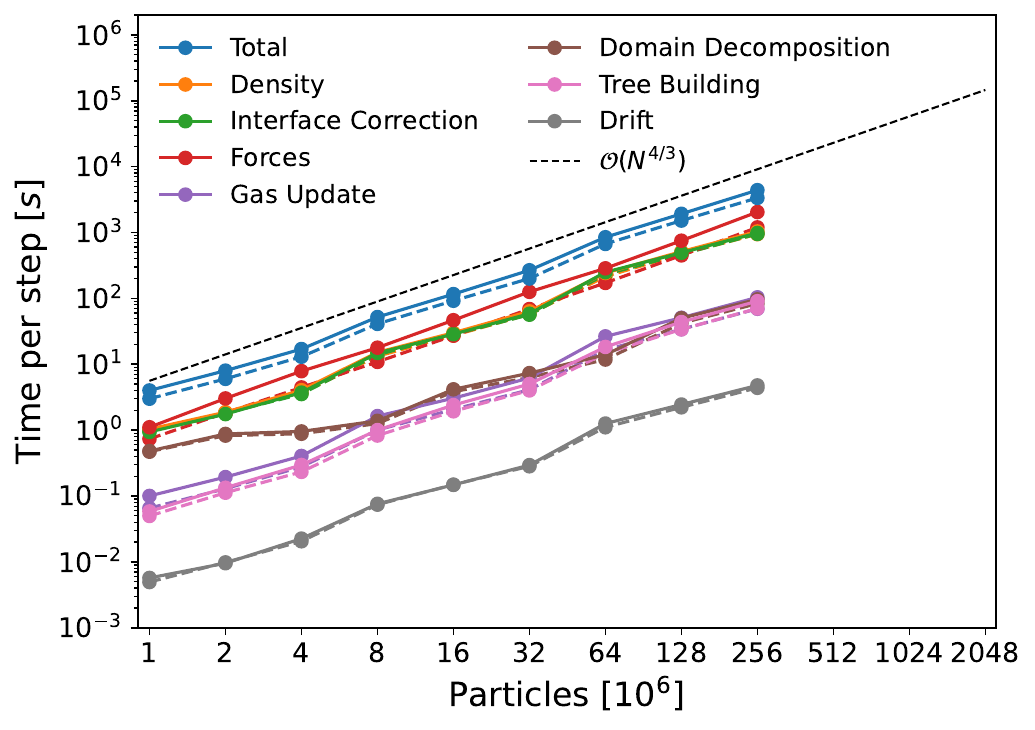}
\includegraphics[width=0.5\linewidth]{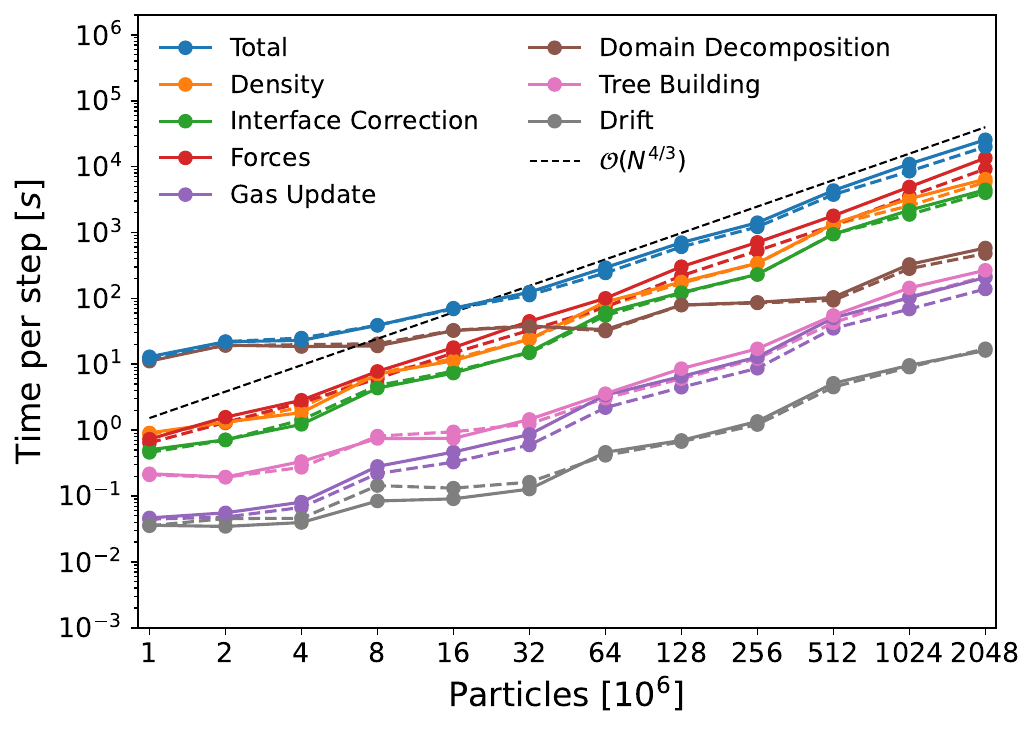}\includegraphics[width=0.5\linewidth]{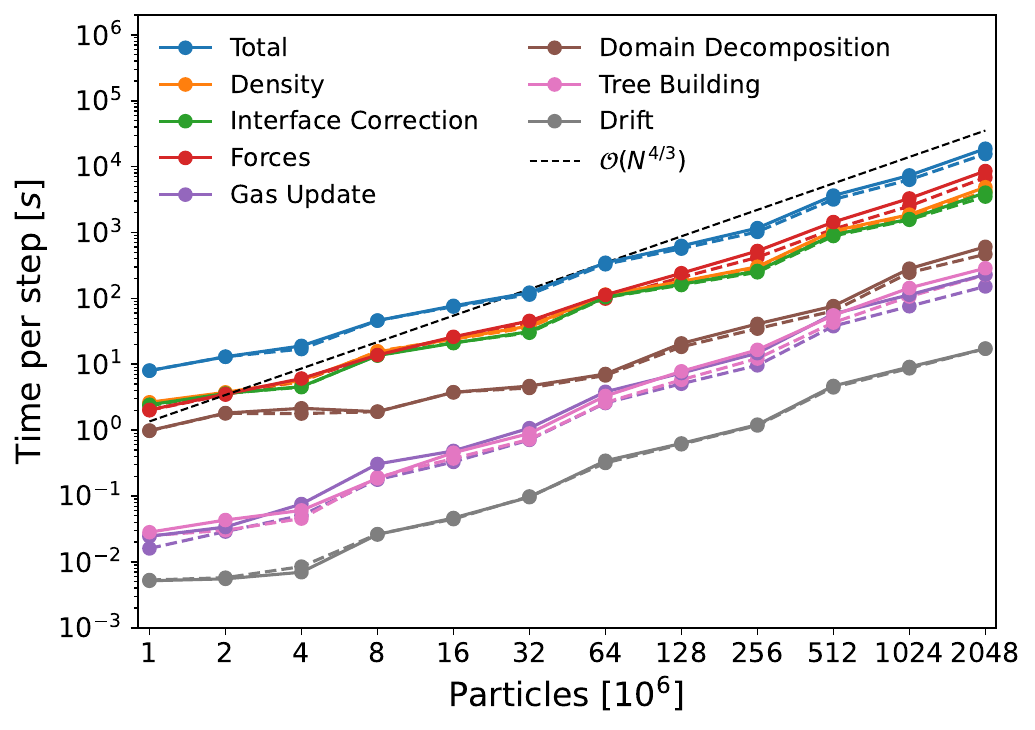}
\caption{Time per step as a function of particle number for all ALPS configurations considered, with (solid lines) and without (dashed lines) the shear strength model enabled. Each panel shows a breakdown of the contributions from the individual computational operations. Results are shown for the 12-thread configuration (top left), the 36-thread configuration (top right), the single-rank 288-thread configuration (bottom left), and the hybrid configuration with 32 MPI ranks and 9 worker threads per rank (bottom right). Each panel includes a reference line indicating the expected $\mathcal{O}\bigl(N^{4/3}\bigr)$ scaling, positioned to highlight the resolution range over which the measured performance follows the asymptotic scaling. Deviations from the reference scaling at low resolutions illustrate the increasing influence of domain decomposition and communication overheads.}
\label{fig:scaling_all_four}
\end{figure*}

In Paper~I, we demonstrated the strong performance and favorable scaling behavior of the SPH implementation in \texttt{pkdgrav3}. The shear strength model is integrated into the same computational pathway as the standard SPH scheme and therefore exhibits closely comparable scaling characteristics. The primary performance question addressed here is the extent to which the increased memory footprint associated with the additional strength fields, (184 vs. 120 bytes per particle), together with the extra computations described in Section~\ref{sec:Code_Description}, introduce measurable overhead.

To quantify this cost, we employ the same benchmark problem used in Paper~I, a Mars-sized body consisting of an M-ANEOS iron \citep{stewartEquationStateModel2020a} core and a M-ANEOS forsterite \citep{stewartEquationStateModel2019} mantle, which is sampled with 1 million to 2 billion particles. These models are then evolved for one global step, the size of which is chosen such that the test with 1 million particles needs 8 substeps. Direct performance comparisons, however, are complicated by the fact that the original benchmark system (Piz Daint) has since been decommissioned and replaced by a substantially more powerful architecture (ALPS). Each Piz Daint node consisted of a single 12-core CPU and one NVIDIA Tesla P100 GPU, whereas each ALPS node comprises four NVIDIA Grace Hopper compute modules, each providing 72 ARM CPU cores with 128~GB of memory and one H100 GPU with 96~GB of HBM3 memory. This corresponds to a factor of 24 increase in CPU core count per node, in addition to a substantial increase in per-core performance.

As a consequence, a test using $10^6$ particles cannot be expected to perform efficiently on a full ALPS node, as this would correspond to fewer than \SI{4000}{} particles per CPU thread, well below the threshold of a few \SI{100000}{} particles per thread identified in Paper~I as a lower bound for acceptable performance. To obtain a meaningful performance assessment under these conditions, we therefore compare the results reported for Piz Daint in Paper~I with a range of ALPS configurations, both with and without the shear strength model enabled.

Figure~\ref{fig:scaling_comparison} presents results for runs using 12 CPU cores (closest to the Piz Daint configuration), 36 cores (the optimal configuration at low resolution), 288 cores (full-node utilization), and a hybrid setup with 32 MPI ranks and 9 worker threads each, which yields the best performance at high resolutions. The substantially improved per-core performance of ALPS is clearly visible in the 12-core results (shown in red) when compared to the Piz Daint data (shown in purple). At a resolution of one million particles, the 36-core configuration reaches a time per step of \SI{3}{\second}, compared to \SI{14}{\second} on Piz Daint. The figure also demonstrates that enabling the shear strength model leads to a non-negligible increase in the time per step.

For the 12- and 36-core configurations, the scaling behavior closely follows that observed on Piz Daint, exhibiting the expected $\mathcal{O}\bigl(N^{4/3}\bigr)$ dependence. In contrast, configurations that utilize the full node (288 cores and the 32-rank hybrid setup) show significant deviations from the expected scaling at low resolutions, where domain decomposition overhead becomes dominant and, at the lowest resolutions, communication costs begin to outweigh computational work (see also Figure~\ref{fig:scaling_all_four}).

Figure~\ref{fig:scaling_best} shows the time per step for the best-performing configuration at each resolution, together with a breakdown of the contributions from the different operations. The resolution range between one million and two billion particles can be divided into three regimes, each dominated by a different optimal configuration. Below eight million particles, the 36-thread configuration is fastest. In the intermediate range between 8 and 64 million particles, the 288-thread single-rank configuration yields the lowest time per step. Above 64 million particles, the hybrid configuration using 32 MPI ranks with 9 worker threads each becomes optimal, as its 32 dispatcher threads are better able to saturate the four GPUs. The transition between configurations occurs at slightly lower resolutions than those at which the individual configurations fully enter their asymptotic $\mathcal{O}\bigl(N^{4/3}\bigr)$ scaling regime (see Figure~\ref{fig:scaling_all_four}), resulting in an overall scaling that is marginally better than the nominal expectation.

The relative ordering of the contributions from individual operations is consistent with that observed on Piz Daint: the three tree-walk operations (density, interface correction, and forces) dominate the total step time, while domain decomposition, tree building, and the gas update contribute significantly less, and the drift contributes negligibly. The contribution from domain decomposition varies strongly with configuration, particularly for the 288-core setup, where it becomes dominant at low resolutions.

The increase in total step time due to the inclusion of the shear strength model is significant. Averaged over all runs shown in Figure~\ref{fig:scaling_best}, the increase amounts to \SI{17}{\percent}, with a maximum increase of \SI{33}{\percent} at low resolutions. As expected, the dominant contribution arises from the forces operation, where the shear strength model introduces substantial additional computations. However, increases are also observed in all other operations, most notably in domain decomposition and tree building, where the larger particle memory footprint increases data movement, and in the gas update, where the prediction of the deviatoric stress tensor (Equation~\eqref{eq:Predict_S}) and additional equation-of-state evaluations for the yield strength add computational cost.

Running simulations with only a few million particles on only a subset of the available resources, for example, using 36 out of 288 CPU cores and a single GPU on a full ALPS node, does not make efficient use of the hardware. For such simulations, we therefore recommend running multiple simulations concurrently on a single node using a workload manager such as HyperQueue \citep{beranekHyperQueueEfficientErgonomic2024}, which is the approach used to obtain the results presented in Sections~\ref{sec:Cliff_collapse} and~\ref{sec:Catastrophic_disruption}. This makes \texttt{pkdgrav3} particularly well suited for large parameter studies at moderate resolutions, enabling efficient utilization of a small number of nodes to execute many simulations concurrently.

\section{Conclusions}\label{sec:Conclusions}
We present the implementation of a shear strength model within the \texttt{pkdgrav3} SPH framework, following \citet{benzImpactSimulationsFracture1994,benzSimulationsBrittleSolids1995, collinsModelingDamageDeformation2004, jutziSPHCalculationsAsteroid2015}, extending the code to model solid and granular materials alongside the previously available hydrodynamic formulation. By leveraging the existing SIMD- and GPU-optimized infrastructure, the strength model introduces moderate computational overhead while preserving the scalability and performance characteristics of the original SPH implementation. The approach incorporates the evolution of the stress tensor, accounting for plasticity and thermal softening, directly into the SPH equations of motion and internal energy evolution, enabling physically realistic simulations of solid-body behavior in a planetary and small-body context.

Validation against a laboratory cliff collapse experiment shows that the model accurately reproduces the observed deposit profile, capturing both the overall shape and the characteristic flow behavior. This agreement demonstrates the high fidelity of the implementation and builds confidence in its predictive capability for more complex scenarios. In catastrophic disruption simulations, we find that material strength significantly increases the disruption threshold $Q_{RD}^*$ and the associated entropy production for small bodies, while the influence of strength diminishes with increasing mass and converges toward the strengthless fluid limit at planetary scales. These results highlight the critical role of material rheology in shaping collision outcomes, particularly for bodies at or below Mars scale, and reinforce the importance of incorporating strength for intermediate-sized objects where self-gravity and internal material strength are comparable.

Furthermore, simulations of low-velocity collisions confirm that shear strength plays a key role in preserving post-impact shapes, such as bi-lobed or elongated remnants, which would otherwise relax toward overly simplified spherical geometries in a purely fluid treatment. While a fully realistic representation of small-body interiors would require additional physics such as damage evolution and porosity, this result demonstrates how shear strength alone can influence remnant morphology and provides a foundation for future high resolution modeling of contact binaries and other irregular objects observed in the solar system. We also confirm that applying the correction tensor to the velocity gradient is essential when shear strength is included, as it reduces otherwise severe errors in angular momentum conservation from percent-level errors to below $10^{-3}$. Although conservation remains slightly worse than in the strength-free case, the correction enables physically accurate simulations of rotating systems. 

We would like to point out that this seems to be the first study that combines recent improvements of SPH for modeling planetary scale collisions \citep{reinhardtNumericalAspectsGiant2017,ruiz-bonillaDealingDensityDiscontinuities2022} with shear strength. The issue addressed by the entropy-conserving scheme is also present in the material strength formulation, and therefore it improves the accuracy of the treatment of entropy, but at the expense of \emph{explicit} energy conservation. We also find that the density correction is essential for reproducing the cliff-collapse experiment, as the standard density estimate (see Equation~\eqref{eq:Density_estimate}) suffers from spurious density smoothing at vacuum interfaces \citep[see][]{reinhardtNumericalAspectsGiant2017}. However, a detailed comparison and validation of different SPH flavors with material strength is beyond the scope of this study. Explicit validation across a broader range of scenarios would be required to verify that these modifications do not lead to unexpected behavior, and will be addressed in future work.

Performance benchmarks indicate that the shear strength model scales efficiently across a wide range of resolutions and computing architectures, with an average increase in time per step of approximately \SI{17}{\percent}. The additional computational cost remains manageable even for simulations exceeding hundreds of millions of particles, enabling unprecedented high-resolution studies of strength-dominated impacts without compromising throughput.

Several extensions of the strength model are already planned. In particular, the addition of explicit cohesive strength, including tensile failure and fracture criteria, together with coupling to damage accumulation and rate-dependent rheologies, would improve the modeling of monolithic materials and enable more accurate simulations of fragmentation, crack-driven disruption, and progressive weakening under repeated loading \citep{meloshDynamicFragmentationImpacts1992,benzSimulationsBrittleSolids1995,collinsModelingDamageDeformation2004,jutziModelingAsteroidCollisions2015}. Incorporating porosity \citep{jutziNumericalSimulationsImpacts2008,wunnemannStrainbasedPorosityModel2006} and crush-curve models \citep{collinsImprovementsaPorous2011,jutziModelingAsteroidCollisions2015} would then allow a more realistic treatment of rubble-pile and highly porous bodies, where compaction and irreversible pore collapse strongly influence impact outcomes and shock propagation. Together, these developments would broaden the applicability of \texttt{pkdgrav3} across the full spectrum of solid-body regimes, from loosely bound granular aggregates to fractured, cohesive planetesimals and differentiated planetary crusts.

\begin{acknowledgments}
We thank the anonymous reviewer for exceptionally thoughtful suggestions and detailed comments that substantially improved the quality and clarity of this paper. This work has been carried out within the framework of the National Centre of Competence in Research PlanetS supported by the Swiss National Science Foundation under grants 51NF40\_182901 and 51NF40\_205606. The authors acknowledge the financial support of the SNSF. TM acknowledges support from the University of Zurich through a Candoc grant. CR and MJ acknowledge support from the Swiss National Science Foundation (project number 200021\_207359). We acknowledge access to Eiger.Alps at the Swiss National Supercomputing Centre, Switzerland under the University of Zurich's share with the project ID UZH4. This work was supported by a grant from the Swiss National Supercomputing Centre (CSCS) under project IDs S1285 and LP87 on Piz Daint and Daint.Alps.
\end{acknowledgments}

\begin{contribution}
T.M. implemented the shear strength module in \texttt{pkdgrav3}, performed the validation and benchmark runs, and was responsible for writing and submitting the manuscript. C.R. assisted T.M. in the implementation, secured the computational resources for the validation runs, and edited the manuscript. M.J. provided expertise on details of the shear strength implementation and edited the manuscript. D.P. is the primary maintainer of the \texttt{pkdgrav3} repository, assisted T.M. with the implementation, and contributed to manuscript editing. J.S. is the original developer of \texttt{pkdgrav1/2/3}, obtained funding to support T.M., and edited the manuscript.
\end{contribution}

%
\facilities{Swiss National Supercomputing Centre (Piz Daint, Eiger.Alps, Daint.Alps)}

\software{pkdgrav3 \citep{potterPKDGRAV3TrillionParticle2017},
          ballic \citep{reinhardtNumericalAspectsGiant2017},
          eoslib \citep{meierEOSlib2021,meierANEOSmaterial2021},
          tipsy \citep{n-bodyshopTIPSYCodeDisplay2011},
          skid \citep{n-bodyshopSKIDFindingGravitationally2011},
          numpy \citep{harrisArrayProgrammingNumPy2020},
          scipy \citep{virtanenSciPy10Fundamental2020},
          matplotlib \citep{hunterMatplotlib2DGraphics2007},
          GNU parallel \citep{tangeGNUParallelCommandline2011},
          HyperQueue \citep{beranekHyperQueueEfficientErgonomic2024}
          }


\appendix
\section{Catastrophic disruption simulation details}\label{sec:appendix:QRDstar_tables}

This appendix summarizes the simulation parameters used in the catastrophic disruption calculations presented in Section~\ref{sec:Catastrophic_disruption}. For each value of $R_{C1}$, a bisection search was performed to determine the value of $Q_{R}$ that yields a bound mass of \SI{50}{\percent} of the total colliding mass. For simulations with constant mass ratio $\gamma$, the bisection was performed in logarithmic impact velocity space, $\log v$, while for simulations with constant impact velocity the bisection was performed in $\gamma$. The iteration was terminated once the relative interval in $Q_R$ between the upper and lower bounds fell below \SI{2}{\percent}, and $Q_{RD}^*$ was obtained by linear interpolation in $Q_{R}$ between the final bracketing values. Tables~\ref{tab:QRDstar_parameters_gamma_0.01}~-~\ref{tab:QRDstar_parameters_v_5} list the corresponding parameters for all data points shown in Section~\ref{sec:Catastrophic_disruption}. Each table entry corresponds to a single simulation for the specified parameter combination; no ensemble averaging over multiple realizations was performed, and all results are based on individual realizations.

\begin{table*}[ht!]
\centering
\begin{tabular}{|c|c|c|c|c|c|c|}
\toprule
& & & \multicolumn{2}{c|}{Fluid Impact velocity $v$} & \multicolumn{2}{c|}{Strength Impact velocity $v$}\\
$R_{C1}$ & Target Mass & Impactor Mass & Low & High & Low & High\\
\midrule
\SI{1.00e5}{\meter} & \SI{6.94e-7}{\Mearth} & \SI{6.90e-9}{\Mearth} &  &  & \SI{4.73}{\kilo\meter\per\second} & \SI{4.78}{\kilo\meter\per\second}\\
\SI{2.15e5}{\meter} & \SI{6.94e-6}{\Mearth} & \SI{6.90e-8}{\Mearth} & \SI{3.86}{\kilo\meter\per\second} & \SI{3.92}{\kilo\meter\per\second} & \SI{6.94}{\kilo\meter\per\second} & \SI{6.98}{\kilo\meter\per\second}\\
\SI{4.64e5}{\meter} & \SI{6.94e-5}{\Mearth} & \SI{6.90e-7}{\Mearth} & \SI{6.26}{\kilo\meter\per\second} & \SI{6.37}{\kilo\meter\per\second} & \SI{10.00}{\kilo\meter\per\second} & \SI{10.17}{\kilo\meter\per\second}\\
\SI{1.00e6}{\meter} & \SI{6.94e-4}{\Mearth} & \SI{6.90e-6}{\Mearth} & \SI{13.93}{\kilo\meter\per\second} & \SI{14.07}{\kilo\meter\per\second} & \SI{16.37}{\kilo\meter\per\second} & \SI{16.55}{\kilo\meter\per\second}\\
\SI{2.15e6}{\meter} & \SI{6.94e-3}{\Mearth} & \SI{6.90e-5}{\Mearth} & \SI{26.51}{\kilo\meter\per\second} & \SI{26.90}{\kilo\meter\per\second} & \SI{29.53}{\kilo\meter\per\second} & \SI{29.96}{\kilo\meter\per\second}\\
\SI{4.64e6}{\meter} & \SI{6.94e-2}{\Mearth} & \SI{6.90e-4}{\Mearth} & \SI{54.25}{\kilo\meter\per\second} & \SI{55.23}{\kilo\meter\per\second} & \SI{56.23}{\kilo\meter\per\second} & \SI{57.11}{\kilo\meter\per\second}\\
\SI{1.00e7}{\meter} & \SI{0.694}{\Mearth} & \SI{6.90e-3}{\Mearth} & \SI{94.75}{\kilo\meter\per\second} & \SI{96.47}{\kilo\meter\per\second} & \SI{94.48}{\kilo\meter\per\second} & \SI{94.75}{\kilo\meter\per\second}\\
\SI{2.15e7}{\meter} & \SI{6.94}{\Mearth} & \SI{6.90e-2}{\Mearth} & \SI{161.19}{\kilo\meter\per\second} & \SI{162.53}{\kilo\meter\per\second} & \SI{159.63}{\kilo\meter\per\second} & \SI{161.38}{\kilo\meter\per\second}\\
\SI{4.64e7}{\meter} & \SI{69.4}{\Mearth} & \SI{0.690}{\Mearth} & \SI{371.80}{\kilo\meter\per\second} & \SI{378.55}{\kilo\meter\per\second} & \SI{371.80}{\kilo\meter\per\second} & \SI{378.55}{\kilo\meter\per\second}\\
\bottomrule
\end{tabular}
\caption{Impact parameters used in the determination of $Q_{RD}^*$ for $\gamma = 0.01$.}
\label{tab:QRDstar_parameters_gamma_0.01}
\end{table*}

\begin{table*}[ht!]
\centering
\begin{tabular}{|c|c|c|c|c|c|c|}
\toprule
& & & \multicolumn{2}{c|}{Fluid Impact velocity $v$} & \multicolumn{2}{c|}{Strength Impact velocity $v$}\\
$R_{C1}$ & Target Mass & Impactor Mass & Low & High & Low & High\\
\midrule
\SI{4.64e5}{\meter} & \SI{6.38e-5}{\Mearth} & \SI{6.38e-6}{\Mearth} & & & \SI{3.59}{\kilo\meter\per\second} & \SI{3.61}{\kilo\meter\per\second}\\
\SI{1.00e6}{\meter} & \SI{6.38e-4}{\Mearth} & \SI{6.38e-5}{\Mearth} & \SI{4.22}{\kilo\meter\per\second} & \SI{4.24}{\kilo\meter\per\second} & \SI{5.20}{\kilo\meter\per\second} & \SI{5.23}{\kilo\meter\per\second}\\
\SI{2.15e6}{\meter} & \SI{6.38e-3}{\Mearth} & \SI{6.38e-4}{\Mearth} & \SI{6.73}{\kilo\meter\per\second} & \SI{6.85}{\kilo\meter\per\second} & \SI{8.08}{\kilo\meter\per\second} & \SI{8.20}{\kilo\meter\per\second}\\
\SI{4.64e6}{\meter} & \SI{6.38e-2}{\Mearth} & \SI{6.38e-3}{\Mearth} & \SI{16.24}{\kilo\meter\per\second} & \SI{16.25}{\kilo\meter\per\second} & \SI{16.51}{\kilo\meter\per\second} & \SI{16.55}{\kilo\meter\per\second}\\
\SI{1.00e7}{\meter} & \SI{0.638}{\Mearth} & \SI{6.38e-2}{\Mearth} & \SI{34.01}{\kilo\meter\per\second} & \SI{34.60}{\kilo\meter\per\second} & \SI{33.98}{\kilo\meter\per\second} & \SI{34.00}{\kilo\meter\per\second}\\
\SI{2.15e7}{\meter} & \SI{6.38}{\Mearth} & \SI{0.638}{\Mearth} & \SI{68.19}{\kilo\meter\per\second} & \SI{68.54}{\kilo\meter\per\second} & \SI{68.18}{\kilo\meter\per\second} & \SI{68.54}{\kilo\meter\per\second}\\
\SI{4.64e7}{\meter} & \SI{63.8}{\Mearth} & \SI{6.38}{\Mearth} & \SI{171.54}{\kilo\meter\per\second} & \SI{172.09}{\kilo\meter\per\second} & \SI{171.80}{\kilo\meter\per\second} & \SI{174.66}{\kilo\meter\per\second}\\
\bottomrule
\end{tabular}
\caption{Impact parameters used in the determination of $Q_{RD}^*$ for $\gamma = 0.1$.}
\label{tab:QRDstar_parameters_gamma_0.1}
\end{table*}

\begin{table*}[ht!]
\centering
\begin{tabular}{|c|c|c|c|c|c|c|}
\toprule
& & & \multicolumn{2}{c|}{Fluid Impact velocity $v$} & \multicolumn{2}{c|}{Strength Impact velocity $v$}\\
$R_{C1}$ & Target Mass & Impactor Mass & Low & High & Low & High\\
\midrule
\SI{1.00e6}{\meter} & \SI{3.51e-4}{\Mearth} & \SI{3.51e-4}{\Mearth} &  &  & \SI{3.15}{\kilo\meter\per\second} & \SI{3.16}{\kilo\meter\per\second}\\
\SI{2.15e6}{\meter} & \SI{3.51e-3}{\Mearth} & \SI{3.51e-3}{\Mearth} & \SI{4.22}{\kilo\meter\per\second} & \SI{4.29}{\kilo\meter\per\second} & \SI{5.05}{\kilo\meter\per\second} & \SI{5.14}{\kilo\meter\per\second}\\
\SI{4.64e6}{\meter} & \SI{3.51e-2}{\Mearth} & \SI{3.51e-2}{\Mearth} & \SI{8.71}{\kilo\meter\per\second} & \SI{8.82}{\kilo\meter\per\second} & \SI{8.98}{\kilo\meter\per\second} & \SI{9.14}{\kilo\meter\per\second}\\
\SI{1.00e7}{\meter} & \SI{0.351}{\Mearth} & \SI{0.351}{\Mearth} & \SI{19.46}{\kilo\meter\per\second} & \SI{19.81}{\kilo\meter\per\second} & \SI{19.46}{\kilo\meter\per\second} & \SI{19.81}{\kilo\meter\per\second}\\
\SI{2.15e7}{\meter} & \SI{3.51}{\Mearth} & \SI{3.51}{\Mearth} & \SI{37.18}{\kilo\meter\per\second} & \SI{37.86}{\kilo\meter\per\second} & \SI{37.18}{\kilo\meter\per\second} & \SI{37.86}{\kilo\meter\per\second}\\
\SI{4.64e7}{\meter} & \SI{35.1}{\Mearth} & \SI{35.1}{\Mearth} & \SI{81.72}{\kilo\meter\per\second} & \SI{82.05}{\kilo\meter\per\second} & \SI{81.72}{\kilo\meter\per\second} & \SI{82.05}{\kilo\meter\per\second}\\
\bottomrule
\end{tabular}
\caption{Impact parameters used in the determination of $Q_{RD}^*$ for $\gamma = 1.0$.}
\label{tab:QRDstar_parameters_gamma_1}
\end{table*}

\begin{table*}[ht!]
\centering
\begin{tabular}{|c|c|c|c|c|c|c|}
\toprule
& & & \multicolumn{2}{c|}{Fluid Impactor Mass} & \multicolumn{2}{c|}{Strength Impactor Mass}\\
$R_{C1}$ & Target Mass & Velocity & Low & High & Low & High\\
\midrule
\SI{4.64e3}{\meter} & \SI{7.01e-11}{\Mearth} & \SI{5.00}{\kilo\meter\per\second} & \SI{3.82e-15}{\Mearth} & \SI{3.87e-15}{\Mearth} & \SI{2.06e-14}{\Mearth} & \SI{2.10e-14}{\Mearth}\\
\SI{1.00e4}{\meter} & \SI{7.01e-10}{\Mearth} & \SI{5.00}{\kilo\meter\per\second} & \SI{9.69e-14}{\Mearth} & \SI{9.76e-14}{\Mearth} & \SI{4.91e-13}{\Mearth} & \SI{4.98e-13}{\Mearth}\\
\SI{2.15e4}{\meter} & \SI{7.01e-9}{\Mearth} & \SI{5.00}{\kilo\meter\per\second} & \SI{2.56e-12}{\Mearth} & \SI{2.61e-12}{\Mearth} & \SI{1.16e-11}{\Mearth} & \SI{1.18e-11}{\Mearth}\\
\SI{4.64e4}{\meter} & \SI{7.01e-8}{\Mearth} & \SI{5.00}{\kilo\meter\per\second} & \SI{7.01e-11}{\Mearth} & \SI{7.05e-11}{\Mearth} & \SI{2.78e-10}{\Mearth} & \SI{2.80e-10}{\Mearth}\\
\SI{1.00e5}{\meter} & \SI{7.01e-7}{\Mearth} & \SI{5.00}{\kilo\meter\per\second} & \SI{1.69e-9}{\Mearth} & \SI{1.72e-9}{\Mearth} & \SI{6.38e-9}{\Mearth} & \SI{6.41e-9}{\Mearth}\\
\SI{2.15e5}{\meter} & \SI{7.01e-6}{\Mearth} & \SI{5.00}{\kilo\meter\per\second} & \SI{4.24e-8}{\Mearth} & \SI{4.30e-8}{\Mearth} & \SI{1.36e-7}{\Mearth} & \SI{1.39e-7}{\Mearth}\\
\SI{4.64e5}{\meter} & \SI{7.01e-5}{\Mearth} & \SI{5.00}{\kilo\meter\per\second} & \SI{1.25e-6}{\Mearth} & \SI{1.27e-6}{\Mearth} & \SI{3.20e-6}{\Mearth} & \SI{3.24e-6}{\Mearth}\\
\SI{1.00e6}{\meter} & \SI{7.01e-4}{\Mearth} & \SI{5.00}{\kilo\meter\per\second} & \SI{4.40e-5}{\Mearth} & \SI{4.47e-5}{\Mearth} & \SI{8.55e-5}{\Mearth} & \SI{8.70e-5}{\Mearth}\\
\bottomrule
\end{tabular}
\caption{Impact parameters used in the determination of $Q_{RD}^*$ for $v=\SI{5}{\kilo\meter\per\second}$.}
\label{tab:QRDstar_parameters_v_5}
\end{table*}


\bibliography{main}{}

@article{ballantyneInvestigatingFeasibilityImpactinduced2023,
  title = {Investigating the Feasibility of an Impact-Induced {{Martian Dichotomy}}},
  author = {Ballantyne, Harry A. and Jutzi, Martin and Golabek, Gregor J. and Mishra, Lokesh and Cheng, Kar Wai and Rozel, Antoine B. and Tackley, Paul J.},
  year = 2023,
  month = mar,
  journal = {Icarus},
  volume = {392},
  pages = {115395},
  issn = {0019-1035},
  doi = {10.1016/j.icarus.2022.115395},
  url = {https://www.sciencedirect.com/science/article/pii/S0019103522004870},
  urldate = {2023-01-19},
  langid = {english}
}

@article{ballantyneSputnikPlanitiaImpactor2024,
  title = {Sputnik {{Planitia}} as an Impactor Remnant Indicative of an Ancient Rocky Mascon in an Oceanless {{Pluto}}},
  author = {Ballantyne, Harry A. and Asphaug, Erik and Denton, C. Adeene and Emsenhuber, Alexandre and Jutzi, Martin},
  year = 2024,
  month = jun,
  journal = {Nature Astronomy},
  volume = {8},
  number = {6},
  pages = {748--755},
  publisher = {Nature Publishing Group},
  issn = {2397-3366},
  doi = {10.1038/s41550-024-02248-1},
  url = {https://www.nature.com/articles/s41550-024-02248-1},
  urldate = {2025-01-30},
  copyright = {2024 The Author(s)},
  langid = {english}
}

@article{benzCatastrophicDisruptionsRevisited1999,
  title = {Catastrophic {{Disruptions Revisited}}},
  author = {Benz, Willy and Asphaug, Erik},
  year = 1999,
  month = nov,
  journal = {Icarus},
  volume = {142},
  number = {1},
  pages = {5--20},
  issn = {0019-1035},
  doi = {10.1006/icar.1999.6204},
  url = {http://www.sciencedirect.com/science/article/pii/S0019103599962048},
  urldate = {2020-10-22},
  langid = {english}
}

@article{benzImpactSimulationsFracture1994,
  title = {Impact {{Simulations}} with {{Fracture}}. {{I}}. {{Method}} and {{Tests}}},
  author = {Benz, W. and Asphaug, E.},
  year = 1994,
  month = jan,
  journal = {Icarus},
  volume = {107},
  number = {1},
  pages = {98--116},
  issn = {0019-1035},
  doi = {10.1006/icar.1994.1009},
  url = {https://www.sciencedirect.com/science/article/pii/S0019103584710098},
  urldate = {2023-01-19},
  langid = {english}
}

@article{benzOriginMoonSingleimpact1986,
  title = {The Origin of the Moon and the Single-Impact Hypothesis {{I}}},
  author = {Benz, W. and Slattery, W. L. and Cameron, A. G. W.},
  year = 1986,
  month = jun,
  journal = {Icarus},
  volume = {66},
  number = {3},
  pages = {515--535},
  issn = {0019-1035},
  doi = {10.1016/0019-1035(86)90088-6},
  url = {http://www.sciencedirect.com/science/article/pii/0019103586900886},
  urldate = {2020-07-03},
  langid = {english}
}

@article{benzSimulationsBrittleSolids1995,
  title = {Simulations of Brittle Solids Using Smooth Particle Hydrodynamics},
  author = {Benz, W. and Asphaug, E.},
  year = 1995,
  month = may,
  journal = {Computer Physics Communications},
  series = {Particle {{Simulation Methods}}},
  volume = {87},
  number = {1},
  pages = {253--265},
  issn = {0010-4655},
  doi = {10.1016/0010-4655(94)00176-3},
  url = {https://www.sciencedirect.com/science/article/pii/0010465594001763},
  urldate = {2022-09-15},
  langid = {english}
}

@article{beranekHyperQueueEfficientErgonomic2024,
  title = {{{HyperQueue}}: {{Efficient}} and Ergonomic Task Graphs on {{HPC}} Clusters},
  shorttitle = {{{HyperQueue}}},
  author = {Ber{\'a}nek, Jakub and B{\"o}hm, Ada and Palermo, Gianluca and Martinovi{\v c}, Jan and Jans{\'i}k, Branislav},
  year = 2024,
  month = sep,
  journal = {SoftwareX},
  volume = {27},
  pages = {101814},
  issn = {2352-7110},
  doi = {10.1016/j.softx.2024.101814},
  url = {https://www.sciencedirect.com/science/article/pii/S2352711024001857},
  urldate = {2026-02-13}
}

@article{bonetVariationalMomentumPreservation1999,
  title = {Variational and Momentum Preservation Aspects of {{Smooth Particle Hydrodynamic}} Formulations},
  author = {Bonet, J. and Lok, T. -S. L.},
  year = 1999,
  month = nov,
  journal = {Computer Methods in Applied Mechanics and Engineering},
  volume = {180},
  number = {1},
  pages = {97--115},
  issn = {0045-7825},
  doi = {10.1016/S0045-7825(99)00051-1},
  url = {https://www.sciencedirect.com/science/article/pii/S0045782599000511},
  urldate = {2025-11-12}
}

@article{bussmannPossibilityGiantImpact2025,
  title = {The Possibility of a Giant Impact on {{Venus}}},
  author = {Bussmann, Mirco and Reinhardt, Christian and Gillmann, Cedric and Meier, Thomas and Stadel, Joachim and Tackley, Paul and Helled, Ravit},
  year = 2025,
  month = oct,
  journal = {Astronomy \& Astrophysics},
  volume = {702},
  pages = {A106},
  publisher = {EDP Sciences},
  issn = {0004-6361, 1432-0746},
  doi = {10.1051/0004-6361/202555802},
  url = {https://www.aanda.org/articles/aa/abs/2025/10/aa55802-25/aa55802-25.html},
  urldate = {2025-10-14},
  copyright = {\copyright{} The Authors 2025},
  langid = {english}
}

@article{cambioniFormationAsteroid162026,
  title = {Formation of {{Asteroid}} (16) {{Psyche}} by a {{Giant Impact}}},
  author = {Cambioni, Saverio and Weiss, Benjamin P. and Baijal, Namya and Melikyan, Robert and Asphaug, Erik and Biersteker, John B. and Binzel, Richard P. and Bottke, William F. and Courville, Samuel W. and {Elkins-Tanton}, Linda T. and Lawrence, David J. and Merayo, Jos{\'e} M. G. and Raymond, Carol A. and Wieczorek, Mark A. and Zuber, Maria T.},
  year = 2026,
  month = jan,
  journal = {Journal of Geophysical Research (Planets)},
  volume = {131},
  pages = {e2025JE009317},
  publisher = {Wiley},
  issn = {0148-0227},
  doi = {10.1029/2025JE009317},
  url = {https://ui.adsabs.harvard.edu/abs/2026JGRE..13109317C},
  urldate = {2026-02-25}
}

@article{canupDynamicsLunarFormation2004,
  title = {Dynamics of {{Lunar Formation}}},
  author = {Canup, Robin M.},
  year = 2004,
  journal = {Annual Review of Astronomy and Astrophysics},
  volume = {42},
  number = {1},
  pages = {441--475},
  doi = {10.1146/annurev.astro.41.082201.113457},
  url = {https://www.annualreviews.org/doi/abs/10.1146/annurev.astro.41.082201.113457},
  urldate = {2023-03-30}
}

@article{canupOriginMoonGiant2001,
  title = {Origin of the {{Moon}} in a Giant Impact near the End of the {{Earth}}'s Formation},
  author = {Canup, Robin M. and Asphaug, Erik},
  year = 2001,
  month = aug,
  journal = {Nature},
  volume = {412},
  number = {6848},
  pages = {708--712},
  publisher = {Nature Publishing Group},
  issn = {1476-4687},
  doi = {10.1038/35089010},
  url = {https://www.nature.com/articles/35089010},
  urldate = {2023-02-16},
  copyright = {2001 Macmillan Magazines Ltd.},
  langid = {english}
}

@article{chauFormingMercuryGiant2018,
  title = {Forming {{Mercury}} by {{Giant Impacts}}},
  author = {Chau, Alice and Reinhardt, Christian and Helled, Ravit and Stadel, Joachim},
  year = 2018,
  month = sep,
  journal = {The Astrophysical Journal},
  volume = {865},
  number = {1},
  pages = {35},
  publisher = {The American Astronomical Society},
  issn = {0004-637X},
  doi = {10.3847/1538-4357/aad8b0},
  url = {https://dx.doi.org/10.3847/1538-4357/aad8b0},
  urldate = {2023-02-16},
  langid = {english}
}

@article{chenImprovementTensileInstability1999,
  title = {An Improvement for Tensile Instability in Smoothed Particle Hydrodynamics},
  author = {Chen, J. K. and Beraun, J. E. and Jih, C. J.},
  year = 1999,
  month = may,
  journal = {Computational Mechanics},
  volume = {23},
  number = {4},
  pages = {279--287},
  issn = {1432-0924},
  doi = {10.1007/s004660050409},
  url = {https://doi.org/10.1007/s004660050409},
  urldate = {2025-11-12},
  langid = {english}
}

@article{collinsImprovementsaPorous2011,
  title = {Improvements to the {\emph{{$\epsilon$}}}-{\emph{{$\alpha$}}} Porous Compaction Model for Simulating Impacts into High-Porosity Solar System Objects},
  author = {Collins, G. S. and Melosh, H. J. and W{\"u}nnemann, K.},
  year = 2011,
  month = jun,
  journal = {International Journal of Impact Engineering},
  series = {Hypervelocity {{Impact}} Selected Papers from the 2010 {{Symposium}}},
  volume = {38},
  number = {6},
  pages = {434--439},
  issn = {0734-743X},
  doi = {10.1016/j.ijimpeng.2010.10.013},
  url = {https://www.sciencedirect.com/science/article/pii/S0734743X10001594},
  urldate = {2025-01-07}
}

@article{collinsModelingDamageDeformation2004,
  title = {Modeling Damage and Deformation in Impact Simulations},
  author = {Collins, Gareth S. and Melosh, H. Jay and Ivanov, Boris A.},
  year = 2004,
  journal = {Meteoritics \& Planetary Science},
  volume = {39},
  number = {2},
  pages = {217--231},
  issn = {1945-5100},
  doi = {10.1111/j.1945-5100.2004.tb00337.x},
  url = {https://onlinelibrary.wiley.com/doi/abs/10.1111/j.1945-5100.2004.tb00337.x},
  urldate = {2023-01-19},
  langid = {english}
}

@article{cukMakingMoonFastSpinning2012,
  title = {Making the {{Moon}} from a {{Fast-Spinning Earth}}: {{A Giant Impact Followed}} by {{Resonant Despinning}}},
  shorttitle = {Making the {{Moon}} from a {{Fast-Spinning Earth}}},
  author = {{\'C}uk, Matija and Stewart, Sarah T.},
  year = 2012,
  month = nov,
  journal = {Science},
  volume = {338},
  number = {6110},
  pages = {1047--1052},
  publisher = {American Association for the Advancement of Science},
  issn = {0036-8075, 1095-9203},
  doi = {10.1126/science.1225542},
  url = {https://science.sciencemag.org/content/338/6110/1047},
  urldate = {2020-05-19},
  chapter = {Research Article},
  copyright = {Copyright \copyright{} 2012, American Association for the Advancement of Science},
  langid = {english},
  pmid = {23076099}
}

@article{dehnenImprovingConvergenceSmoothed2012,
  title = {Improving Convergence in Smoothed Particle Hydrodynamics Simulations without Pairing Instability},
  author = {Dehnen, Walter and Aly, Hossam},
  year = 2012,
  month = sep,
  journal = {Monthly Notices of the Royal Astronomical Society},
  volume = {425},
  number = {2},
  pages = {1068--1082},
  issn = {0035-8711},
  doi = {10.1111/j.1365-2966.2012.21439.x},
  url = {https://doi.org/10.1111/j.1365-2966.2012.21439.x},
  urldate = {2023-10-25}
}

@article{dentonCaptureAncientCharon2025,
  title = {Capture of an Ancient {{Charon}} around {{Pluto}}},
  author = {Denton, C. Adeene and Asphaug, Erik and Emsenhuber, Alexandre and Melikyan, Robert},
  year = 2025,
  month = jan,
  journal = {Nature Geoscience},
  volume = {18},
  number = {1},
  pages = {37--43},
  publisher = {Nature Publishing Group},
  issn = {1752-0908},
  doi = {10.1038/s41561-024-01612-0},
  url = {https://www.nature.com/articles/s41561-024-01612-0},
  urldate = {2025-08-12},
  copyright = {2025 The Author(s), under exclusive licence to Springer Nature Limited},
  langid = {english}
}

@article{druckerSoilMechanicsPlastic1952,
  title = {Soil {{Mechanics}} and {{Plastic Analysis}} or {{Limit Design}}},
  author = {Drucker, D. C. and Prager, W.},
  year = 1952,
  journal = {Quarterly of Applied Mathematics},
  volume = {10},
  number = {2},
  eprint = {43633942},
  eprinttype = {jstor},
  pages = {157--165},
  publisher = {Brown University},
  issn = {0033-569X},
  url = {https://www.jstor.org/stable/43633942},
  urldate = {2026-02-11}
}

@article{emsenhuberSPHCalculationsMarsscale2018,
  title = {{{SPH}} Calculations of {{Mars-scale}} Collisions: {{The}} Role of the Equation of State, Material Rheologies, and Numerical Effects},
  shorttitle = {{{SPH}} Calculations of {{Mars-scale}} Collisions},
  author = {Emsenhuber, Alexandre and Jutzi, Martin and Benz, Willy},
  year = 2018,
  month = feb,
  journal = {Icarus},
  volume = {301},
  pages = {247--257},
  issn = {00191035},
  doi = {10.1016/j.icarus.2017.09.017},
  url = {https://linkinghub.elsevier.com/retrieve/pii/S0019103517302397},
  urldate = {2020-05-05},
  langid = {english}
}

@article{gendaResolutionDependenceDisruptive2015,
  title = {Resolution Dependence of Disruptive Collisions between Planetesimals in the Gravity Regime},
  author = {Genda, Hidenori and Fujita, Tomoaki and Kobayashi, Hiroshi and Tanaka, Hidekazu and Abe, Yutaka},
  year = 2015,
  month = dec,
  journal = {Icarus},
  volume = {262},
  pages = {58--66},
  issn = {00191035},
  doi = {10.1016/j.icarus.2015.08.029},
  url = {https://linkinghub.elsevier.com/retrieve/pii/S0019103515003735},
  urldate = {2020-05-05},
  langid = {english}
}

@article{golabekCouplingSPHThermochemical2018,
  title = {Coupling {{SPH}} and Thermochemical Models of Planets: {{Methodology}} and Example of a {{Mars-sized}} Body},
  shorttitle = {Coupling {{SPH}} and Thermochemical Models of Planets},
  author = {Golabek, G. J. and Emsenhuber, A. and Jutzi, M. and Asphaug, E. I. and Gerya, T. V.},
  year = 2018,
  month = feb,
  journal = {Icarus},
  volume = {301},
  pages = {235--246},
  issn = {0019-1035},
  doi = {10.1016/j.icarus.2017.10.003},
  url = {https://www.sciencedirect.com/science/article/pii/S0019103517302385},
  urldate = {2026-02-06}
}

@article{gonzalez-cataldoMeltingCurveSiO22016,
  title = {Melting Curve of {{SiO2}} at Multimegabar Pressures: Implications for Gas Giants and Super-{{Earths}}},
  shorttitle = {Melting Curve of {{SiO2}} at Multimegabar Pressures},
  author = {{Gonz{\'a}lez-Cataldo}, Felipe and Davis, Sergio and Guti{\'e}rrez, Gonzalo},
  year = 2016,
  month = may,
  journal = {Scientific Reports},
  volume = {6},
  number = {1},
  pages = {26537},
  publisher = {Nature Publishing Group},
  issn = {2045-2322},
  doi = {10.1038/srep26537},
  url = {https://www.nature.com/articles/srep26537},
  urldate = {2026-02-09},
  copyright = {2016 The Author(s)},
  langid = {english}
}

@article{gradyContinuumModellingExplosive1980,
  title = {Continuum Modelling of Explosive Fracture in Oil Shale},
  author = {Grady, D. E. and Kipp, M. E.},
  year = 1980,
  month = jun,
  journal = {International Journal of Rock Mechanics and Mining Sciences \& Geomechanics Abstracts},
  volume = {17},
  number = {3},
  pages = {147--157},
  issn = {0148-9062},
  doi = {10.1016/0148-9062(80)91361-3},
  url = {https://www.sciencedirect.com/science/article/pii/0148906280913613},
  urldate = {2026-02-11}
}

@article{harrisArrayProgrammingNumPy2020,
  title = {Array Programming with {{NumPy}}},
  author = {Harris, Charles R. and Millman, K. Jarrod and {van der Walt}, St{\'e}fan J. and Gommers, Ralf and Virtanen, Pauli and Cournapeau, David and Wieser, Eric and Taylor, Julian and Berg, Sebastian and Smith, Nathaniel J. and Kern, Robert and Picus, Matti and Hoyer, Stephan and {van Kerkwijk}, Marten H. and Brett, Matthew and Haldane, Allan and {del R{\'i}o}, Jaime Fern{\'a}ndez and Wiebe, Mark and Peterson, Pearu and {G{\'e}rard-Marchant}, Pierre and Sheppard, Kevin and Reddy, Tyler and Weckesser, Warren and Abbasi, Hameer and Gohlke, Christoph and Oliphant, Travis E.},
  year = 2020,
  month = sep,
  journal = {Nature},
  volume = {585},
  number = {7825},
  pages = {357--362},
  publisher = {Nature Publishing Group},
  issn = {1476-4687},
  doi = {10.1038/s41586-020-2649-2},
  url = {https://www.nature.com/articles/s41586-020-2649-2},
  urldate = {2023-02-16},
  copyright = {2020 The Author(s)},
  langid = {english}
}

@article{holsappleModelingGranularMaterial2013,
  title = {Modeling Granular Material Flows: {{The}} Angle of Repose, Fluidization and the Cliff Collapse Problem},
  shorttitle = {Modeling Granular Material Flows},
  author = {Holsapple, Keith A.},
  year = 2013,
  month = jul,
  journal = {Planetary and Space Science},
  volume = {82--83},
  pages = {11--26},
  issn = {0032-0633},
  doi = {10.1016/j.pss.2013.03.001},
  url = {https://www.sciencedirect.com/science/article/pii/S0032063313000573},
  urldate = {2025-01-07}
}

@article{hopkinsGeneralClassLagrangian2013,
  title = {A General Class of {{Lagrangian}} Smoothed Particle Hydrodynamics Methods and Implications for Fluid Mixing Problems},
  author = {Hopkins, Philip F.},
  year = 2013,
  month = feb,
  journal = {Monthly Notices of the Royal Astronomical Society},
  volume = {428},
  number = {4},
  pages = {2840--2856},
  issn = {0035-8711, 1365-2966},
  doi = {10.1093/mnras/sts210},
  url = {http://academic.oup.com/mnras/article/428/4/2840/992468/A-general-class-of-Lagrangian-smoothed-particle},
  urldate = {2021-03-09},
  langid = {english}
}

@article{hunterMatplotlib2DGraphics2007,
  title = {Matplotlib: {{A 2D Graphics Environment}}},
  shorttitle = {Matplotlib},
  author = {Hunter, John D.},
  year = 2007,
  month = may,
  journal = {Computing in Science Engineering},
  volume = {9},
  number = {3},
  pages = {90--95},
  issn = {1558-366X},
  doi = {10.1109/MCSE.2007.55},
  url = {https://ieeexplore.ieee.org/document/4160265}
}

@article{jopConstitutiveLawDense2006,
  title = {A Constitutive Law for Dense Granular Flows},
  author = {Jop, Pierre and Forterre, Yo{\"e}l and Pouliquen, Olivier},
  year = 2006,
  month = jun,
  journal = {Nature},
  volume = {441},
  number = {7094},
  pages = {727--730},
  publisher = {Nature Publishing Group},
  issn = {1476-4687},
  doi = {10.1038/nature04801},
  url = {https://www.nature.com/articles/nature04801},
  urldate = {2026-01-29},
  copyright = {2006 Springer Nature Limited},
  langid = {english}
}

@article{jutziFragmentPropertiesCatastrophic2010,
  title = {Fragment Properties at the Catastrophic Disruption Threshold: {{The}} Effect of the Parent Body's Internal Structure},
  shorttitle = {Fragment Properties at the Catastrophic Disruption Threshold},
  author = {Jutzi, Martin and Michel, Patrick and Benz, Willy and Richardson, Derek C.},
  year = 2010,
  month = may,
  journal = {Icarus},
  volume = {207},
  number = {1},
  pages = {54--65},
  issn = {0019-1035},
  doi = {10.1016/j.icarus.2009.11.016},
  url = {https://www.sciencedirect.com/science/article/pii/S0019103509004564},
  urldate = {2026-02-06}
}

@article{jutziModelingAsteroidCollisions2015,
  title = {Modeling Asteroid Collisions and Impact Processes},
  author = {Jutzi, Martin and Holsapple, Keith and W{\"u}nneman, Kai and Michel, Patrick},
  year = 2015,
  journal = {arXiv:1502.01844 [astro-ph]},
  eprint = {1502.01844},
  primaryclass = {astro-ph},
  doi = {10.2458/azu_uapress_9780816532131-ch035},
  url = {http://arxiv.org/abs/1502.01844},
  urldate = {2020-05-05},
  archiveprefix = {arXiv}
}

@article{jutziNumericalSimulationsImpacts2008,
  title = {Numerical Simulations of Impacts Involving Porous Bodies: {{I}}. {{Implementing}} Sub-Resolution Porosity in a {{3D SPH}} Hydrocode},
  shorttitle = {Numerical Simulations of Impacts Involving Porous Bodies},
  author = {Jutzi, Martin and Benz, Willy and Michel, Patrick},
  year = 2008,
  month = nov,
  journal = {Icarus},
  volume = {198},
  number = {1},
  pages = {242--255},
  issn = {0019-1035},
  doi = {10.1016/j.icarus.2008.06.013},
  url = {https://www.sciencedirect.com/science/article/pii/S0019103508002558},
  urldate = {2023-01-19},
  langid = {english}
}

@article{jutziShapeStructureCometary2015,
  title = {The Shape and Structure of Cometary Nuclei as a Result of Low-Velocity Accretion},
  author = {Jutzi, M. and Asphaug, E.},
  year = 2015,
  month = jun,
  journal = {Science},
  volume = {348},
  number = {6241},
  pages = {1355--1358},
  publisher = {American Association for the Advancement of Science},
  doi = {10.1126/science.aaa4747},
  url = {https://www.science.org/doi/10.1126/science.aaa4747},
  urldate = {2026-02-09}
}

@article{jutziShapeStructureSmall2019,
  title = {The Shape and Structure of Small Asteroids as a Result of Sub-Catastrophic Collisions},
  author = {Jutzi, Martin},
  year = 2019,
  month = nov,
  journal = {Planetary and Space Science},
  volume = {177},
  pages = {104695},
  issn = {0032-0633},
  doi = {10.1016/j.pss.2019.07.009},
  url = {https://www.sciencedirect.com/science/article/pii/S0032063319300868},
  urldate = {2026-02-11}
}

@article{jutziSPHCalculationsAsteroid2015,
  title = {{{SPH}} Calculations of Asteroid Disruptions: {{The}} Role of Pressure Dependent Failure Models},
  shorttitle = {{{SPH}} Calculations of Asteroid Disruptions},
  author = {Jutzi, Martin},
  year = 2015,
  month = mar,
  journal = {Planetary and Space Science},
  series = {{{VIII Workshop}} on {{Catastrophic Disruption}} in the {{Solar System}}},
  volume = {107},
  pages = {3--9},
  issn = {0032-0633},
  doi = {10.1016/j.pss.2014.09.012},
  url = {https://www.sciencedirect.com/science/article/pii/S0032063314002931},
  urldate = {2023-01-19},
  langid = {english}
}

@article{kegerreisImmediateOriginMoon2022,
  title = {Immediate {{Origin}} of the {{Moon}} as a {{Post-impact Satellite}}},
  author = {Kegerreis, J. A. and {Ruiz-Bonilla}, S. and Eke, V. R. and Massey, R. J. and Sandnes, T. D. and Teodoro, L. F. A.},
  year = 2022,
  month = oct,
  journal = {The Astrophysical Journal Letters},
  volume = {937},
  number = {2},
  pages = {L40},
  publisher = {The American Astronomical Society},
  issn = {2041-8205},
  doi = {10.3847/2041-8213/ac8d96},
  url = {https://dx.doi.org/10.3847/2041-8213/ac8d96},
  urldate = {2023-02-16},
  langid = {english}
}

@article{kiuchiImpactExperimentsGranular2023,
  title = {Impact Experiments on Granular Materials under Low Gravity: {{Effects}} of Cohesive Strength, Internal Friction, and Porosity of Particle Layers on Crater Size},
  shorttitle = {Impact Experiments on Granular Materials under Low Gravity},
  author = {Kiuchi, Masato and Okamoto, Takaya and Nagaashi, Yuuya and Yamaguchi, Yukari and Hasegawa, Sunao and Nakamura, Akiko M.},
  year = 2023,
  month = nov,
  journal = {Icarus},
  volume = {404},
  pages = {115685},
  issn = {0019-1035},
  doi = {10.1016/j.icarus.2023.115685},
  url = {https://www.sciencedirect.com/science/article/pii/S0019103523002622},
  urldate = {2026-02-11}
}

@article{lajeunesseGranularSlumpingHorizontal2005,
  title = {Granular Slumping on a Horizontal Surface},
  author = {Lajeunesse, E. and Monnier, J. B. and Homsy, G. M.},
  year = 2005,
  month = oct,
  journal = {Physics of Fluids},
  volume = {17},
  number = {10},
  pages = {103302},
  issn = {1070-6631, 1089-7666},
  doi = {10.1063/1.2087687},
  url = {https://pubs.aip.org/pof/article/17/10/103302/985425/Granular-slumping-on-a-horizontal-surface},
  urldate = {2026-01-29},
  langid = {english}
}

@article{leinhardtCollisionsGravitydominatedBodies2012,
  title = {Collisions between {{Gravity-dominated Bodies}}. {{I}}. {{Outcome Regimes}} and {{Scaling Laws}}},
  author = {Leinhardt, Zo{\"e} M. and Stewart, Sarah T.},
  year = 2012,
  month = jan,
  journal = {The Astrophysical Journal},
  volume = {745},
  pages = {79},
  doi = {10.1088/0004-637X/745/1/79},
  url = {http://adsabs.harvard.edu/abs/2012ApJ...745...79L},
  urldate = {2020-09-17}
}

@article{leleuPeculiarShapesSaturns2018,
  title = {The Peculiar Shapes of {{Saturn}}'s Small Inner Moons as Evidence of Mergers of Similar-Sized Moonlets},
  author = {Leleu, A. and Jutzi, M. and Rubin, M.},
  year = 2018,
  month = jul,
  journal = {Nature Astronomy},
  volume = {2},
  number = {7},
  pages = {555--561},
  publisher = {Nature Publishing Group},
  issn = {2397-3366},
  doi = {10.1038/s41550-018-0471-7},
  url = {https://www.nature.com/articles/s41550-018-0471-7},
  urldate = {2026-02-11},
  copyright = {2018 The Author(s)},
  langid = {english}
}

@article{lineweaverPotatoRadiusLower2010,
  title = {The {{Potato Radius}}: A {{Lower Minimum Size}} for {{Dwarf Planets}}},
  shorttitle = {The {{Potato Radius}}},
  author = {Lineweaver, Charles H. and Norman, Marc},
  year = 2010,
  month = apr,
  journal = {arXiv:1004.1091 [astro-ph, physics:physics]},
  eprint = {1004.1091},
  primaryclass = {astro-ph, physics:physics},
  url = {http://arxiv.org/abs/1004.1091},
  urldate = {2020-07-08},
  archiveprefix = {arXiv}
}

@article{lucyNumericalApproachTesting1977,
  title = {A Numerical Approach to the Testing of the Fission Hypothesis},
  author = {Lucy, L. B.},
  year = 1977,
  month = dec,
  journal = {The Astronomical Journal},
  volume = {82},
  pages = {1013--1024},
  issn = {0004-6256},
  doi = {10.1086/112164},
  url = {http://adsabs.harvard.edu/abs/1977AJ.....82.1013L},
  urldate = {2021-06-04}
}

@article{maindlSPHbasedSimulationMultimaterial2013,
  title = {{{SPH-based}} Simulation of Multi-Material Asteroid Collisions},
  author = {Maindl, T.i. and Sch{\"a}fer, C. and Speith, R. and S{\"u}li, {\'A}. and {Forg{\'a}cs-Dajka}, E. and Dvorak, R.},
  year = 2013,
  journal = {Astronomische Nachrichten},
  volume = {334},
  number = {9},
  pages = {996--999},
  issn = {1521-3994},
  doi = {10.1002/asna.201311979},
  url = {https://onlinelibrary.wiley.com/doi/abs/10.1002/asna.201311979},
  urldate = {2026-02-24},
  langid = {english}
}

@article{marohnicConstrainingFinalMerger2021,
  title = {Constraining the Final Merger of Contact Binary (486958) {{Arrokoth}} with Soft-Sphere Discrete Element Simulations},
  author = {Marohnic, J. C. and Richardson, D. C. and McKinnon, W. B. and Agrusa, H. F. and DeMartini, J. V. and Cheng, A. F. and Stern, S. A. and Olkin, C. B. and Weaver, H. A. and Spencer, J. R.},
  year = 2021,
  month = mar,
  journal = {Icarus},
  series = {Pluto {{System}}, {{Kuiper Belt}}, and {{Kuiper Belt Objects}}},
  volume = {356},
  pages = {113824},
  issn = {0019-1035},
  doi = {10.1016/j.icarus.2020.113824},
  url = {https://www.sciencedirect.com/science/article/pii/S0019103520302062},
  urldate = {2026-03-17}
}

@article{mcglaunCTHThreedimensionalShock1990,
  title = {{{CTH}}: {{A}} Three-Dimensional Shock Wave Physics Code},
  shorttitle = {{{CTH}}},
  author = {McGlaun, J. M. and Thompson, S. L. and Elrick, M. G.},
  year = 1990,
  month = jan,
  journal = {International Journal of Impact Engineering},
  volume = {10},
  number = {1},
  pages = {351--360},
  issn = {0734-743X},
  doi = {10.1016/0734-743X(90)90071-3},
  url = {https://www.sciencedirect.com/science/article/pii/0734743X90900713},
  urldate = {2026-02-25}
}

@misc{meierANEOSmaterial2021,
  title = {{{ANEOSmaterial}}},
  shorttitle = {{{ANEOSmaterial}}},
  author = {Meier, Thomas and Reinhardt, Christian},
  year = 2021,
  month = apr,
  publisher = {Zenodo},
  doi = {10.5281/zenodo.4662606},
  url = {https://zenodo.org/record/4662606},
  urldate = {2021-04-05}
}

@misc{meierEOSlib2021,
  title = {{{EOSlib}}},
  shorttitle = {{{EOSlib}}},
  author = {Meier, Thomas and Reinhardt, Christian},
  year = 2021,
  month = apr,
  publisher = {Zenodo},
  doi = {10.5281/zenodo.4662637},
  url = {https://zenodo.org/record/4662637},
  urldate = {2021-04-05}
}

@article{meierEOSResolutionConspiracy2021,
  title = {The {{EOS}}/Resolution Conspiracy: Convergence in Proto-Planetary Collision Simulations},
  shorttitle = {The {{EOS}}/Resolution Conspiracy},
  author = {Meier, Thomas and Reinhardt, Christian and Stadel, Joachim G.},
  year = 2021,
  month = jun,
  journal = {Monthly Notices of the Royal Astronomical Society},
  volume = {505},
  number = {2},
  pages = {1806--1816},
  issn = {0035-8711, 1365-2966},
  doi = {10.1093/mnras/stab1441},
  url = {https://academic.oup.com/mnras/article/505/2/1806/6279686},
  urldate = {2021-09-06},
  langid = {english}
}

@article{meierOriginJupitersFuzzy2025,
  title = {On the {{Origin}} of {{Jupiter}}'s {{Fuzzy Core}}: {{Constraints}} from {{N-body}}, {{Impact}}, and {{Evolution Simulations}}},
  shorttitle = {On the {{Origin}} of {{Jupiter}}'s {{Fuzzy Core}}},
  author = {Meier, Thomas and Reinhardt, Christian and Shibata, Sho and M{\"u}ller, Simon and Stadel, Joachim and Helled, Ravit},
  year = 2025,
  month = jul,
  journal = {The Astrophysical Journal},
  volume = {988},
  number = {1},
  pages = {7},
  publisher = {The American Astronomical Society},
  issn = {0004-637X},
  doi = {10.3847/1538-4357/addbe6},
  url = {https://dx.doi.org/10.3847/1538-4357/addbe6},
  urldate = {2025-08-07},
  langid = {english}
}

@article{meierSmoothedParticleHydrodynamics2026,
  title = {Smoothed {{Particle Hydrodynamics}} in Pkdgrav3 for {{Shock Physics Simulations}}. {{I}}. {{Hydrodynamics}}},
  author = {Meier, Thomas and Potter, Douglas and Reinhardt, Christian and Stadel, Joachim},
  year = 2026,
  month = mar,
  journal = {The Astrophysical Journal},
  volume = {1000},
  number = {2},
  pages = {266},
  publisher = {The American Astronomical Society},
  issn = {0004-637X},
  doi = {10.3847/1538-4357/ae4e29},
  url = {https://doi.org/10.3847/1538-4357/ae4e29},
  urldate = {2026-03-27},
  langid = {english}
}

@article{meierSystematicSurveyMoonforming2024,
  title = {A {{Systematic Survey}} of {{Moon-forming Giant Impacts}}. {{II}}. {{Rotating Bodies}}},
  author = {Meier, Thomas and Reinhardt, Christian and Timpe, Miles and Stadel, Joachim and Moore, Ben},
  year = 2024,
  month = dec,
  journal = {The Astrophysical Journal},
  volume = {978},
  number = {1},
  pages = {11},
  publisher = {The American Astronomical Society},
  issn = {0004-637X},
  doi = {10.3847/1538-4357/ad9248},
  url = {https://dx.doi.org/10.3847/1538-4357/ad9248},
  urldate = {2025-01-07},
  langid = {english}
}

@article{meloshDynamicFragmentationImpacts1992,
  title = {Dynamic Fragmentation in Impacts: {{Hydrocode}} Simulation of Laboratory Impacts},
  shorttitle = {Dynamic Fragmentation in Impacts},
  author = {Melosh, H. J. and Ryan, E. V. and Asphaug, E.},
  year = 1992,
  journal = {Journal of Geophysical Research: Planets},
  volume = {97},
  number = {E9},
  pages = {14735--14759},
  issn = {2156-2202},
  doi = {10.1029/92JE01632},
  url = {https://onlinelibrary.wiley.com/doi/abs/10.1029/92JE01632},
  urldate = {2025-01-07},
  copyright = {Copyright 1992 by the American Geophysical Union.},
  langid = {english}
}

@article{meloshHydrocodeEquationState2007,
  title = {A Hydrocode Equation of State for {{SiO}} {\textsubscript{2}}},
  author = {Melosh, H. J.},
  year = 2007,
  month = dec,
  journal = {Meteoritics \& Planetary Science},
  volume = {42},
  number = {12},
  pages = {2079--2098},
  issn = {10869379, 19455100},
  doi = {10.1111/j.1945-5100.2007.tb01009.x},
  url = {http://doi.wiley.com/10.1111/j.1945-5100.2007.tb01009.x},
  urldate = {2020-05-05},
  langid = {english}
}

@article{michaelowenCompatiblyDifferencedTotal2014,
  title = {A Compatibly Differenced Total Energy Conserving Form of {{SPH}}},
  author = {Michael Owen, J.},
  year = 2014,
  journal = {International Journal for Numerical Methods in Fluids},
  volume = {75},
  number = {11},
  pages = {749--774},
  issn = {1097-0363},
  doi = {10.1002/fld.3912},
  url = {https://onlinelibrary.wiley.com/doi/abs/10.1002/fld.3912},
  urldate = {2026-02-24},
  langid = {english}
}

@article{monaghanShockSimulationParticle1983,
  title = {Shock Simulation by the Particle Method {{SPH}}},
  author = {Monaghan, J. J and Gingold, R. A},
  year = 1983,
  month = nov,
  journal = {Journal of Computational Physics},
  volume = {52},
  number = {2},
  pages = {374--389},
  issn = {0021-9991},
  doi = {10.1016/0021-9991(83)90036-0},
  url = {http://www.sciencedirect.com/science/article/pii/0021999183900360},
  urldate = {2020-06-26},
  langid = {english}
}

@article{monaghanSmoothedParticleHydrodynamics1992,
  title = {Smoothed Particle Hydrodynamics},
  author = {Monaghan, J. J.},
  year = 1992,
  journal = {Annual Review of Astronomy and Astrophysics},
  volume = {30},
  pages = {543--574},
  issn = {0066-4146},
  doi = {10.1146/annurev.aa.30.090192.002551},
  url = {http://adsabs.harvard.edu/abs/1992ARA%26A..30..543M},
  urldate = {2020-07-08}
}

@article{n-bodyshopSKIDFindingGravitationally2011,
  title = {{{SKID}}: {{Finding Gravitationally Bound Groups}} in {{N-body Simulations}}},
  shorttitle = {{{SKID}}},
  author = {{N-Body Shop}},
  year = 2011,
  month = feb,
  journal = {Astrophysics Source Code Library},
  pages = {ascl:1102.020},
  url = {https://ui.adsabs.harvard.edu/abs/2011ascl.soft02020N},
  urldate = {2023-08-28}
}

@article{n-bodyshopTIPSYCodeDisplay2011,
  title = {{{TIPSY}}: {{Code}} for {{Display}} and {{Analysis}} of {{N-body Simulations}}},
  shorttitle = {{{TIPSY}}},
  author = {{N-Body Shop}},
  year = 2011,
  month = nov,
  journal = {Astrophysics Source Code Library},
  pages = {ascl:1111.015},
  url = {https://ui.adsabs.harvard.edu/abs/2011ascl.soft11015N},
  urldate = {2025-01-31}
}

@article{nakamuraImpactCrateringPorous2017,
  title = {Impact Cratering on Porous Targets in the Strength Regime},
  author = {Nakamura, Akiko M.},
  year = 2017,
  month = dec,
  journal = {Planetary and Space Science},
  series = {Special {{Issue}}: {{Cosmic Dust IX}}},
  volume = {149},
  pages = {5--13},
  issn = {0032-0633},
  doi = {10.1016/j.pss.2017.09.001},
  url = {https://www.sciencedirect.com/science/article/pii/S003206331730051X},
  urldate = {2026-02-11}
}

@article{ohnakaShearFailureStrength1995,
  title = {A Shear Failure Strength Law of Rock in the Brittle-Plastic Transition Regime},
  author = {Ohnaka, Mitiyasu},
  year = 1995,
  journal = {Geophysical Research Letters},
  volume = {22},
  number = {1},
  pages = {25--28},
  issn = {1944-8007},
  doi = {10.1029/94GL02791},
  url = {https://onlinelibrary.wiley.com/doi/abs/10.1029/94GL02791},
  urldate = {2026-02-03},
  copyright = {Copyright 1995 by the American Geophysical Union.},
  langid = {english}
}

@misc{potterDpotterPkdgrav3V3512026,
  title = {Dpotter/Pkdgrav3: V3.5.1},
  shorttitle = {Dpotter/Pkdgrav3},
  author = {Potter, Douglas and Stadel, Joachim Gerhard and Meier, Thomas},
  year = 2026,
  month = feb,
  publisher = {Zenodo},
  doi = {10.5281/zenodo.18754678},
  url = {https://zenodo.org/records/18754678},
  urldate = {2026-04-24}
}

@article{potterPKDGRAV3TrillionParticle2017,
  title = {{{PKDGRAV3}}: Beyond Trillion Particle Cosmological Simulations for the next Era of Galaxy Surveys},
  shorttitle = {{{PKDGRAV3}}},
  author = {Potter, Douglas and Stadel, Joachim and Teyssier, Romain},
  year = 2017,
  month = dec,
  journal = {Computational Astrophysics and Cosmology},
  volume = {4},
  number = {1},
  pages = {2},
  issn = {2197-7909},
  doi = {10.1186/s40668-017-0021-1},
  url = {https://comp-astrophys-cosmol.springeropen.com/articles/10.1186/s40668-017-0021-1},
  urldate = {2021-03-15},
  langid = {english}
}

@article{priceSmoothedParticleHydrodynamics2012,
  title = {Smoothed Particle Hydrodynamics and Magnetohydrodynamics},
  author = {Price, Daniel J.},
  year = 2012,
  month = feb,
  journal = {Journal of Computational Physics},
  series = {Special {{Issue}}: {{Computational Plasma Physics}}},
  volume = {231},
  number = {3},
  pages = {759--794},
  issn = {0021-9991},
  doi = {10.1016/j.jcp.2010.12.011},
  url = {https://www.sciencedirect.com/science/article/pii/S0021999110006753},
  urldate = {2021-03-09},
  langid = {english}
}

@article{raducanMultipleMoonletMergers2025,
  title = {Multiple Moonlet Mergers as the Origin of the {{Dinkinesh-Selam}} System},
  author = {Raducan, S. D. and Madeira, G. and Agrusa, H. F. and Merrill, C. C. and Marschall, R. and Ferrari, F. and Wimarsson, J. and Charnoz, S. and Michel, P. and Jutzi, M.},
  year = 2025,
  month = dec,
  journal = {Nature Communications},
  volume = {16},
  number = {1},
  pages = {11033},
  publisher = {Nature Publishing Group},
  issn = {2041-1723},
  doi = {10.1038/s41467-025-66484-3},
  url = {https://www.nature.com/articles/s41467-025-66484-3},
  urldate = {2026-02-24},
  copyright = {2025 The Author(s)},
  langid = {english}
}

@article{raducanPhysicalPropertiesAsteroid2024,
  title = {Physical Properties of Asteroid {{Dimorphos}} as Derived from the {{DART}} Impact},
  author = {Raducan, S. D. and Jutzi, M. and Cheng, A. F. and Zhang, Y. and Barnouin, O. and Collins, G. S. and Daly, R. T. and Davison, T. M. and Ernst, C. M. and Farnham, T. L. and Ferrari, F. and Hirabayashi, M. and Kumamoto, K. M. and Michel, P. and Murdoch, N. and Nakano, R. and Pajola, M. and Rossi, A. and Agrusa, H. F. and Barbee, B. W. and Syal, M. Bruck and Chabot, N. L. and Dotto, E. and Fahnestock, E. G. and Hasselmann, P. H. and Herreros, I. and Ivanovski, S. and Li, J.-Y. and Lucchetti, A. and Luther, R. and Orm{\"o}, J. and Owen, M. and Pravec, P. and Rivkin, A. S. and Robin, C. Q. and S{\'a}nchez, P. and Tusberti, F. and W{\"u}nnemann, K. and Zinzi, A. and Epifani, E. Mazzotta and Manzoni, C. and May, B. H.},
  year = 2024,
  month = apr,
  journal = {Nature Astronomy},
  volume = {8},
  number = {4},
  pages = {445--455},
  publisher = {Nature Publishing Group},
  issn = {2397-3366},
  doi = {10.1038/s41550-024-02200-3},
  url = {https://www.nature.com/articles/s41550-024-02200-3},
  urldate = {2026-03-24},
  copyright = {2024 The Author(s)},
  langid = {english}
}

@article{reinhardtBifurcationHistoryUranus2020,
  title = {Bifurcation in the History of {{Uranus}} and {{Neptune}}: The Role of Giant Impacts},
  shorttitle = {Bifurcation in the History of {{Uranus}} and {{Neptune}}},
  author = {Reinhardt, Christian and Chau, Alice and Stadel, Joachim and Helled, Ravit},
  year = 2020,
  month = mar,
  journal = {Monthly Notices of the Royal Astronomical Society},
  volume = {492},
  number = {4},
  pages = {5336--5353},
  publisher = {Oxford Academic},
  issn = {0035-8711},
  doi = {10.1093/mnras/stz3271},
  url = {https://academic.oup.com/mnras/article/492/4/5336/5637902},
  urldate = {2020-08-10},
  langid = {english}
}

@article{reinhardtFormingIronrichPlanets2022,
  title = {Forming Iron-Rich Planets with Giant Impacts},
  author = {Reinhardt, Christian and Meier, Thomas and Stadel, Joachim G. and Otegi, Jon F and Helled, Ravit},
  year = 2022,
  month = dec,
  journal = {Monthly Notices of the Royal Astronomical Society},
  volume = {517},
  number = {3},
  pages = {3132--3143},
  issn = {0035-8711},
  doi = {10.1093/mnras/stac1853},
  url = {https://doi.org/10.1093/mnras/stac1853},
  urldate = {2023-01-12}
}

@article{reinhardtNumericalAspectsGiant2017,
  title = {Numerical Aspects of {{Giant Impact}} Simulations},
  author = {Reinhardt, Christian and Stadel, Joachim},
  year = 2017,
  month = jan,
  journal = {Monthly Notices of the Royal Astronomical Society},
  volume = {467},
  doi = {10.1093/mnras/stx322},
  url = {http://adsabs.harvard.edu/abs/2017MNRAS.467.4252R}
}

@article{ruiz-bonillaDealingDensityDiscontinuities2022,
  title = {Dealing with Density Discontinuities in Planetary {{SPH}} Simulations},
  author = {{Ruiz-Bonilla}, S and Borrow, J and Eke, V R and Kegerreis, J A and Massey, R J and Sandnes, T D and Teodoro, L F A},
  year = 2022,
  month = may,
  journal = {Monthly Notices of the Royal Astronomical Society},
  volume = {512},
  number = {3},
  pages = {4660--4668},
  issn = {0035-8711},
  doi = {10.1093/mnras/stac857},
  url = {https://doi.org/10.1093/mnras/stac857},
  urldate = {2023-10-25}
}

@article{safronovRelativeSizesLargest1969,
  title = {Relative Sizes of the Largest Bodies during the Accumulation of Planets},
  author = {Safronov, V. S. and Zvjagina, E. V.},
  year = 1969,
  month = jan,
  journal = {Icarus},
  volume = {10},
  number = {1},
  pages = {109--115},
  issn = {0019-1035},
  doi = {10.1016/0019-1035(69)90013-X},
  url = {https://www.sciencedirect.com/science/article/pii/001910356990013X},
  urldate = {2026-02-11}
}

@article{schaferCollisionsEqualsizedIce2007,
  title = {Collisions between Equal-Sized Ice Grain Agglomerates},
  author = {Sch{\"a}fer, C. and Speith, R. and Kley, W.},
  year = 2007,
  month = aug,
  journal = {Astronomy \& Astrophysics},
  volume = {470},
  number = {2},
  pages = {733--739},
  issn = {0004-6361, 1432-0746},
  doi = {10.1051/0004-6361:20077354},
  url = {http://www.aanda.org/10.1051/0004-6361:20077354},
  urldate = {2023-11-15},
  langid = {english}
}

@article{schaferSmoothParticleHydrodynamics2016,
  title = {A Smooth Particle Hydrodynamics Code to Model Collisions between Solid, Self-Gravitating Objects},
  author = {Sch{\"a}fer, C. and Riecker, S. and Maindl, T. I. and Speith, R. and Scherrer, S. and Kley, W.},
  year = 2016,
  month = jun,
  journal = {Astronomy \& Astrophysics},
  volume = {590},
  pages = {A19},
  publisher = {EDP Sciences},
  issn = {0004-6361, 1432-0746},
  doi = {10.1051/0004-6361/201528060},
  url = {https://www.aanda.org/articles/aa/abs/2016/06/aa28060-15/aa28060-15.html},
  urldate = {2020-05-05},
  copyright = {\copyright{} ESO, 2016},
  langid = {english}
}

@article{springelCosmologicalSmoothedParticle2002,
  title = {Cosmological Smoothed Particle Hydrodynamics Simulations: The Entropy Equation},
  shorttitle = {Cosmological Smoothed Particle Hydrodynamics Simulations},
  author = {Springel, Volker and Hernquist, Lars},
  year = 2002,
  month = jul,
  journal = {Monthly Notices of the Royal Astronomical Society},
  volume = {333},
  number = {3},
  pages = {649--664},
  publisher = {Oxford Academic},
  issn = {0035-8711},
  doi = {10.1046/j.1365-8711.2002.05445.x},
  url = {https://academic.oup.com/mnras/article/333/3/649/1002394},
  urldate = {2020-10-29},
  langid = {english}
}

@misc{stewartEquationStateModel2019,
  title = {Equation of {{State Model Forsterite-ANEOS-SLVTv1}}.{{0G1}}: {{Documentation}} and {{Comparisons}}},
  shorttitle = {Equation of {{State Model Forsterite-ANEOS-SLVTv1}}.{{0G1}}},
  author = {Stewart, Sarah T. and Davies, Erik J. and Duncan, Megan S. and Lock, Simon J. and Root, Seth and Townsend, Joshua P. and Kraus, Richard G. and Caracas, Razvan and Jacobsen, Stein B.},
  year = 2019,
  month = oct,
  publisher = {Zenodo},
  doi = {10.5281/zenodo.3478631},
  url = {https://zenodo.org/record/3478631},
  urldate = {2023-08-08}
}

@misc{stewartEquationStateModel2020a,
  title = {Equation of {{State Model Iron ANEOS}}: {{Documentation}} and {{Comparisons}} ({{Version SLVTv0}}.{{2G1}})},
  shorttitle = {Equation of {{State Model Iron ANEOS}}},
  author = {Stewart, Sarah T.},
  year = 2020,
  month = may,
  publisher = {Zenodo},
  doi = {10.5281/zenodo.3866507},
  url = {https://zenodo.org/record/3866507},
  urldate = {2023-08-31}
}

@article{stewartVELOCITYDEPENDENTCATASTROPHICDISRUPTION2009,
  title = {{{VELOCITY-DEPENDENT CATASTROPHIC DISRUPTION CRITERIA FOR PLANETESIMALS}}},
  author = {Stewart, Sarah T. and Leinhardt, Zo{\"e} M.},
  year = 2009,
  month = jan,
  journal = {The Astrophysical Journal},
  volume = {691},
  number = {2},
  pages = {L133},
  publisher = {The American Astronomical Society},
  issn = {0004-637X},
  doi = {10.1088/0004-637X/691/2/L133},
  url = {https://doi.org/10.1088/0004-637X/691/2/L133},
  urldate = {2026-02-06},
  langid = {english}
}

@article{sugiuraHighresolutionSimulationsCatastrophic2020,
  title = {High-Resolution Simulations of Catastrophic Disruptions: {{Resultant}} Shape Distributions},
  shorttitle = {High-Resolution Simulations of Catastrophic Disruptions},
  author = {Sugiura, Keisuke and Kobayashi, Hiroshi and Inutsuka, Shu-ichiro},
  year = 2020,
  month = feb,
  journal = {Planetary and Space Science},
  volume = {181},
  pages = {104807},
  issn = {0032-0633},
  doi = {10.1016/j.pss.2019.104807},
  url = {https://www.sciencedirect.com/science/article/pii/S0032063319300303},
  urldate = {2026-03-30}
}

@article{tangeGNUParallelCommandline2011,
  title = {{{GNU}} Parallel - the Command-Line Power Tool},
  author = {Tange, O.},
  year = 2011,
  month = feb,
  journal = {;login: The USENIX Magazine},
  volume = {36},
  number = {1},
  pages = {42--47},
  address = {Frederiksberg, Denmark},
  url = {http://www.gnu.org/s/parallel}
}

@techreport{thompsonImprovementsCHARTRadiationhydrodynamic1974,
  title = {Improvements in the {{CHART D}} Radiation-Hydrodynamic Code {{III}}: Revised Analytic Equations of State},
  shorttitle = {Improvements in the {{CHART D}} Radiation-Hydrodynamic Code {{III}}},
  author = {Thompson, S. L. and Lauson, H. S.},
  year = 1974,
  number = {SC-RR--71-0714},
  institution = {Sandia Labs.},
  url = {http://inis.iaea.org/Search/search.aspx?orig_q=RN:6209386},
  urldate = {2020-05-12},
  langid = {english}
}

@misc{thompsonMANEOS2019,
  title = {M-{{ANEOS}}},
  author = {Thompson, S. L. and Lauson, H. S. and Melosh, H. J. and Collins, G. S. and Stewart, S. T.},
  year = 2019,
  month = nov,
  publisher = {Zenodo},
  doi = {10.5281/zenodo.3525030},
  url = {https://zenodo.org/record/3525030#.XvxXHygzaUk},
  urldate = {2020-07-01}
}

@article{timpeSystematicSurveyMoonforming2023,
  title = {A {{Systematic Survey}} of {{Moon-forming Giant Impacts}}. {{I}}. {{Nonrotating Bodies}}},
  author = {Timpe, Miles and Reinhardt, Christian and Meier, Thomas and Stadel, Joachim and Moore, Ben},
  year = 2023,
  month = dec,
  journal = {The Astrophysical Journal},
  volume = {959},
  number = {1},
  pages = {38},
  publisher = {The American Astronomical Society},
  issn = {0004-637X},
  doi = {10.3847/1538-4357/acfc40},
  url = {https://dx.doi.org/10.3847/1538-4357/acfc40},
  urldate = {2024-01-05},
  langid = {english}
}

@article{virtanenSciPy10Fundamental2020,
  title = {{{SciPy}} 1.0: Fundamental Algorithms for Scientific Computing in {{Python}}},
  shorttitle = {{{SciPy}} 1.0},
  author = {Virtanen, Pauli and Gommers, Ralf and Oliphant, Travis E. and Haberland, Matt and Reddy, Tyler and Cournapeau, David and Burovski, Evgeni and Peterson, Pearu and Weckesser, Warren and Bright, Jonathan and {van der Walt}, St{\'e}fan J. and Brett, Matthew and Wilson, Joshua and Millman, K. Jarrod and Mayorov, Nikolay and Nelson, Andrew R. J. and Jones, Eric and Kern, Robert and Larson, Eric and Carey, C. J. and Polat, {\.I}lhan and Feng, Yu and Moore, Eric W. and VanderPlas, Jake and Laxalde, Denis and Perktold, Josef and Cimrman, Robert and Henriksen, Ian and Quintero, E. A. and Harris, Charles R. and Archibald, Anne M. and Ribeiro, Ant{\^o}nio H. and Pedregosa, Fabian and {van Mulbregt}, Paul},
  year = 2020,
  month = mar,
  journal = {Nature Methods},
  volume = {17},
  number = {3},
  pages = {261--272},
  publisher = {Nature Publishing Group},
  issn = {1548-7105},
  doi = {10.1038/s41592-019-0686-2},
  url = {https://www.nature.com/articles/s41592-019-0686-2},
  urldate = {2023-02-16},
  copyright = {2020 The Author(s)},
  langid = {english}
}

@article{wimarssonDiverseShapesBinary2025,
  title = {The Diverse Shapes of Binary Asteroid Satellites Born from Sub-Escape-Velocity Moonlet Mergers},
  author = {Wimarsson, John and Ferrari, Fabio and Jutzi, Martin},
  year = 2025,
  month = dec,
  journal = {Astronomy \& Astrophysics},
  volume = {704},
  pages = {A29},
  publisher = {EDP Sciences},
  issn = {0004-6361, 1432-0746},
  doi = {10.1051/0004-6361/202555914},
  url = {https://www.aanda.org/articles/aa/abs/2025/12/aa55914-25/aa55914-25.html},
  urldate = {2026-02-11},
  copyright = {\copyright{} The Authors 2025},
  langid = {english}
}

@article{wunnemannStrainbasedPorosityModel2006,
  title = {A Strain-Based Porosity Model for Use in Hydrocode Simulations of Impacts and Implications for Transient Crater Growth in Porous Targets},
  author = {W{\"u}nnemann, K. and Collins, G. S. and Melosh, H. J.},
  year = 2006,
  month = feb,
  journal = {Icarus},
  volume = {180},
  pages = {514--527},
  issn = {0019-1035},
  doi = {10.1016/j.icarus.2005.10.013},
  url = {https://ui.adsabs.harvard.edu/abs/2006Icar..180..514W},
  urldate = {2025-01-07}
}
\bibliographystyle{aasjournalv7}



\end{document}